\documentclass[journal,onecolumn,12pt]{IEEEtran}
\usepackage{diagbox}
\usepackage{graphicx}
\usepackage{epstopdf}
\usepackage{threeparttable}
\usepackage{amsmath,amssymb}
\usepackage{arydshln}
\usepackage{dsfont}
\usepackage{color}
\usepackage{cases}
\usepackage{bm}
\usepackage{bbding}
\usepackage{amsfonts}
\usepackage[font=footnotesize]{caption}
\usepackage{indentfirst}
\usepackage{cite}
\usepackage{caption}
\usepackage{algorithm}
\usepackage{algpseudocode}
\usepackage{booktabs}
\usepackage{subfigure}
\usepackage{multirow}
\usepackage{amsthm}
\usepackage{setspace}
\newcommand{\sd}{\mathbf}

\usepackage{hyperref}

\theoremstyle{remark}
\newtheorem{remark}{Remark}
\theoremstyle{}
\newtheorem{theorem}{Theorem}
\theoremstyle{}

\newtheorem{lemma}{Lemma}
\theoremstyle{}
\newtheorem{definition}{Definition}
\theoremstyle{remark}
\newtheorem{example}{Example}

\theoremstyle{definition}

\makeatletter
\newcommand{\tabcaption}{\def\@captype{table}\caption}

\makeatletter
 \allowdisplaybreaks[4]

\ifCLASSINFOpdf
\else
\fi

\definecolor{newcolor}{rgb}{0.5,0,1}

\newcommand{\yq}[1]{{\color{black}#1}}

\begin{document}

\title{\yq{Robust, Secure and Private Cache-aided  Scalar Linear Function Retrieval from Distributed System with Blind and Adversarial Servers}}
\author{
Qifa Yan~\IEEEmembership{Member,~IEEE,} Zhengchun Zhou,~\IEEEmembership{Member,~IEEE}, 

and Xiaohu Tang,~\IEEEmembership{Senior Member,~IEEE,}

\thanks{The authors are with the Information Coding \& Transmission Key Lab of Sichuan Province, CSNMT Int. Coop. Res. Centre (MoST), Southwest Jiaotong University, Chengdu 611756, China.   (email: qifayan@swjtu.edu.cn, zzc@swjtu.edu.cn,  xhutang@swjtu.edu.cn)
}
}

\maketitle
\pagestyle{empty}  
\thispagestyle{empty} 
\IEEEpeerreviewmaketitle

\begin{abstract}
\yq{In this work, a distributed server system composed of multiple servers that holds some coded files and multiple users that are interested in retrieving the linear functions of the files is investigated, where the servers are robust, blind and adversarial in the sense that any $J$ servers can together recover all files, while any $I$ colluding servers cannot obtain any information about the files, and  at most $A$ servers maliciously provides erroneous information. In addition, the file library must be secure from a wiretapper who obtains all the signals, and the demands of any subset of users must kept private from the other users and servers, even if they collude.} 
A coding scheme is proposed by incorporating the ideas of Shamir's secret sharing and key superposition into the framework of Placement Delivery Array (PDA), originally proposed to characterize the single-server coded caching system without any security or privacy constraints. It is shown that PDAs associated to Maddah-Ali and Niesen's coded caching scheme results in an achievable memory-storage-communication region, such that the storage size and communication load were optimal to within a multiplicative gap, except for the small memory regime when the number of files was smaller than the number of users.

\end{abstract}
\section{Introduction}
\label{sec:intro}

Coded caching, a technique introduced by Maddah-Ali and Niesen (MAN) in 2014, reduces peak-time communication loads by jointly designing pre-stored cache contents and delivered communication signals~\cite{Maddah-Ali2014fundamental}. The system consists of a single server holding a set of files that connects to multiple cache-aided users through a shared link. It operates in two phases: the placement phase, when the server fills the users' caches without knowledge of their future demands, and the delivery phase, when the server satisfies the users' demands by transmitting coded signals to them, having knowledge of their demands.

For a system with $N$ files and $K$ users, the MAN scheme achieves the optimal load-memory tradeoff among all uncoded placement schemes when $N\geq K$~\cite{Kai2020Index} and for $N<K$ after removing some redundant transmissions~\cite{Yu2019ExactTradeoff}. However, one important problem in the MAN scheme is that the number of packets each file needs to be partitioned into (called subpacketization) grows exponentially with the number of users $K$ \cite{FiniteLength}. In 2017, Yan et al. proposed a \yq{combinatorial} object called a Placement Delivery Array (PDA) to characterize a subclass of coded caching schemes with uncoded placement, which was utilized to design coded caching schemes with low subpacketization schemes \cite{Yan2017PDA,PDA:bipartite}. As a tool, PDA has been widely used in many networks, e.g., device-to-device networks  \cite{PDAD2D}, multi-access networks \cite{MultaccessPDA}, and coded distributed computing \cite{SCC,straglingPDA}. Two constructions of PDA were proposed in that work, which had similar communication loads but with significantly smaller subpacketization compared to the MAN scheme. Recently, it was shown that allowing the users to demand arbitrary linear combinations of the files does not increase the communication load compared to the case of single file retrieval, at least under uncoded placement~\cite{Kai2020LinearFunction}.

During the delivery phase, signals are typically transmitted on a public channel, and users need to share their demands, which raises critical issues of content security and demand privacy in practical systems. In~\cite{Security}, the content of the library must be protected against an external wiretapper who obtains the signals transmitted during the delivery phase. This was achieved by pre-storinging security keys into user caches for the part of the files that were not cached in the MAN scheme.

Specially, the security keys are stored in a structured way such that each user can decode all the multicast signals it needs to decode. In~\cite{Kai2019Private,Kamath2019}, users' demand privacy are guarranteed by adding virtual users, so that the real users can not distinguish if the demands are from real or virtual users. Another technique to guarantee privacy is privacy key proposed in~\cite{Y:D:Privacy}, which allows the users to privately \yq{retrieve} arbitrary scalar linear combination of the files with significantly lower subpacketization. The key idea in~\cite{Y:D:Privacy} is that, each user also privately caches some random linear combination of the parts of the files, which was called privacy keys. The queries are generated by adding the coefficients used to generate the privacy keys to the real demands, so that the users can decode the linear combination of files with the queries and further decode their own demands with the privacy keys. In~\cite{Y:D:SP-LFR}, a key superposition scheme was proposed to guarantee both content security against a wiretapper and demand privacy against colluding users simultaneously. The idea is superposing (i.e., sum together) the security keys and privacy keys such that the load-memory tradeoff in this case is the same as in the setup with only content security guarantees.

The progresses mentioned above have been achieved under a basic setup that involves a single server and multiple users. However, in a dual setup that consists of multiple servers and a single user, the demand privacy for such networks is known as Private Information Retrieval (PIR) \cite{ChorPIR95}, \cite{HusPIRCapacity}. Coding across the distributed servers with Maximum Distance Separable (MDS) codes is a useful technique since it saves storage while allowing node failures or erasures, which commonly occur in practical systems \cite{Dimarkis2011}. The schemes achieving the optimal communication load with almost optimal sub-packetization are proposed in \cite{Jinbao2020,ZhouMDS_PIR}. For multiple-server-multiple-user systems, the techniques from coded caching and PIR were combined in \cite{XiangZhang, XiangZhang02} to guarantee server-side privacy. In \cite{ISITversion,Yan2022JSAC}, the technique of key superposition is integrated into the MDS-coded server system under the PDA framework, further guaranteeing demand privacy against both servers and colluding users, signal security against an eavesdropper, and robust decoding against server failures.

\begin{table*}[!htp]
 \centering
\caption{Works on Cache-Aided Multi-User-Multi-Server System}\label{table1}
 \small{\begin{tabular}{|c|c|c|c|}
\hline
Properties&Cache Aided PIR\cite{XiangZhang,XiangZhang02}&RSP-LFR\cite{ISITversion,Yan2022JSAC}&RSP-LFR-BA(this work)\\\hline
Privacy against colluding users&\XSolidBrush&\Checkmark&\Checkmark\\\hline
Privacy against a single server&\Checkmark&\Checkmark&\Checkmark\\\hline
Privacy against colluding servers&\XSolidBrush&\Checkmark&\Checkmark\\\hline
Content security at the servers&\XSolidBrush&\XSolidBrush&\Checkmark\\\hline
Signal security against eavesdropper&\XSolidBrush&\Checkmark&\Checkmark\\\hline
Robustness against link failures&\XSolidBrush&\Checkmark&\Checkmark\\\hline
\yq{Robustness against adversarial servers}&\XSolidBrush&\XSolidBrush&\Checkmark\\\hline
  \end{tabular}}
\end{table*}

In distributed systems, a new breed of security that is gaining popularity is the \yq{“blind and adversarial servers”}, where servers are untrusted and are required to remain ignorant of the data, \yq{while some servers maliciously provide erroneous information.} This is a crucial aspect for many scenarios, as the data stored at servers is often accessible to the public. 
In this paper, we go beyond the usual requirements of demand privacy, signal security, and robust decoding, and introduce a new criterion for the design of distributed storage at servers. We ensure that the content of the files remains concealed from any subset of $I$ servers, and  the entire file can be retrieved from the contents of any $J$ servers, \yq{in the presence of at most $A$ adversarial servers.} We term this model as “Robust, Secure, and Private Scalar Linear Function Retrieval from Blind \yq{and Adversarial} Servers (RSP-LFR-BA)”.

Our main contributions towards the proposed RSP-LFR-BA model are:
\begin{enumerate}
\item We introduce a method to generate an RSP-LFR-BA scheme from a given PDA. The advantage of utilizing the PDA framework is that RSP-LFR-BA schemes with low subpacketizations can be easily obtained from a range of existing PDA constructions, such as those in~\cite{Yan2017PDA,PDA:bipartite,PDA:a,PDA:b,PDA:c}. The key idea is to combine the Shamir secret sharing and key superposition techniques within the PDA framework in the multi-server-multi-user setup. In this setup, security keys are encoded using a polynomial to align them with the signals in the Shamir secret sharing polynomials \cite{Shamir1979
}. The memory size at each user, the storage size at the servers, and the communication load are characterized using the PDA parameters.

\item Following the proposed method, we derive an achievable region of memory-storage-communication triples based on PDAs that describe the original MAN scheme in~\cite{Maddah-Ali2014fundamental} (MAN-PDAs). Additionally, the achievable storage size and communication load are optimal within a  multiplicative gap, except for the small memory regime when the number of files is smaller than the number of users.
\end{enumerate}
Table \ref{table1} compares the properties of several recent works on cache-aided multi-user-multi-server systems.

The rest of this paper is organized as follows.
Section~\ref{sec:model} gives the formal system model description.
Section~\ref{sec:PDA:example} reviews the PDA framework and gives an illustrative example.
Section~\ref{sec:main} summarizes our main results, where the proof details are deferred to Sections \ref{sec:scheme} and \ref{sec:MAN:PDA}.
Section~\ref{sec:numerical} presents some numerical results, and
Section~\ref{sec:conclusion} concludes the paper.

\paragraph*{Notations} 
In this paper, we use the following conventions:
\begin{itemize}
\item Let $\mathbb{N}^+$ be the set of positive integers. Let $\mathbb{F}_q$ and $\mathbb{F}_q^n$ be the finite field of cardinality  $q$, for some prime power $q$, and the $n$-dimensional vector space over $\mathbb{F}_q$, respectively. 
\item For two integers $m,n$ such that $m\leq n$, we use $[m:n]$ to denote the set of the first positive integers $\{m,\ldots,n\}$; $[1:n]$ is also denoted by $[n]$ for short. 
\item We use $X_{\mathcal{A}}$ to denote the tuple composed of $\{X_i:i\in\mathcal{A}\}$ for the integer set $\mathcal{A}$, where the elements are ordered increasingly, e.g., $X_{[3]}=(X_1,X_2,X_3)$. For variables with two or more indices, e.g., $X_{i,j}$, we use $X_{\mathcal{A},\mathcal{B}}$ to denote the tuple $\{X_{i,j}: i\in\mathcal{A},j\in\mathcal{B} \}$, where the elements are listed in lexicographical order, e.g. $X_{[3],[2]}=(X_{1,1},X_{1,2},X_{2,1},X_{2,2},X_{3,1},X_{3,2})$.
\item  Let $N,I,J\in\mathbb{N}^+$. For a set of $N$ vectors over the finite field $\mathbb{F}_q$ of the same dimension, say, $X_1,\ldots,X_N$, and a given vector $\sd{b}=(b_1,\ldots,b_N)$, we use $X_{\sd{b}}$ to denote the linear combination of $X_1,\ldots,X_N$ with coefficient vector $\sd{b}$, i.e., $X_{\sd{b}}:=\sum_{n=1}^Nb_n\cdot X_n$.

\quad If each vector $X_n$ is composed of $I$ sub-vectors of the same dimension, say $X_n=(X_{n,1},\ldots,X_{n,I})$, we use $X_{\sd{b},i}$ to denote linear combination of the vectors $X_{1,i},\ldots,X_{n,i}$ with coefficient vector $\sd{b}$, i.e., $X_{\sd{b},i}:=\sum_{n=1}^Nb_n\cdot X_{n,i}$. Similarly, if each sub-vector $X_{n,i}$ is composed of $J$ sub-sub-vectors of the same dimension, say $X_{n,i}=(X_{n,i,1},\ldots,X_{n,i,J})$, we use $X_{\sd{b},i,j}$ to denote the linear combination of the sub-sub-vectors $X_{1,i,j},\ldots,X_{N,i,j}$ with coefficient vector $\sd{b}$, i.e., $X_{\sd{b},i,j}:=\sum_{n=1}^Nb_n\cdot X_{n,i,j}$. 

\quad For any $\mathcal{A}\subseteq[I]$, the notation $X_{\sd{b},\mathcal{A}}$ denotes the tuple composed of $\{X_{\sd{b},i}:i\in\mathcal{A}\}$, where the elements are ordered increasingly, e.g., $X_{\sd{b},[3]}=(X_{\sd{b},1},X_{\sd{b},2},X_{\sd{b},3})$. Similarly,  for any $\mathcal{A}\subseteq[I],\mathcal{B}\subseteq[J]$, $X_{\sd{b},\mathcal{A},\mathcal{B}}$ denotes the tuple composed of $\{X_{\sd{b},i,j}:i\in\mathcal{A},j\in\mathcal{B}\}$,  where the elements are listed in lexicographical order, e.g., $X_{\sd{b},[3],[2]}=(X_{\sd{b},1,1},X_{\sd{b},1,2},X_{\sd{b},2,1},X_{\sd{b},2,2},X_{\sd{b},3,1},X_{\sd{b},3,2})$. If $\mathcal{A}$ or $\mathcal{B}$ is of cardinality one, we will abbreviate $X_{\sd{b},\{i\},\mathcal{B}}$ and $X_{\sd{b},\mathcal{A},\{j\}}$ as $X_{\sd{b},i,\mathcal{B}}$ and $X_{\sd{b},\mathcal{A},j}$, respectively. 

\end{itemize}


\section{System Model}\label{sec:model}
\begin{figure}[htb!]
  \centering
  \includegraphics[width=0.7\columnwidth]{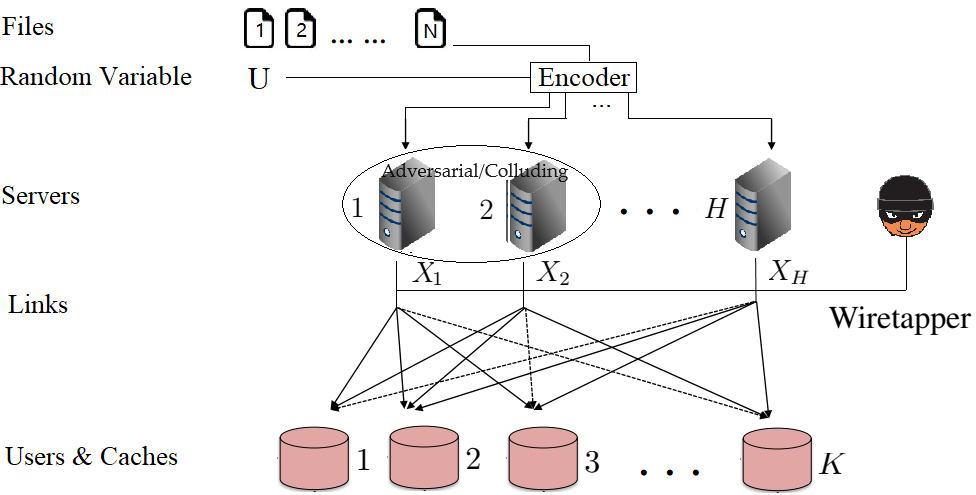}
  \caption{System model}\label{fig:system}
\vspace{-10pt}
\end{figure}

\yq{Let $N,K,H,A,I,J$ be positive integers such that $A\leq I\leq J\leq H$ and $I+2A< J$}. 
A $(N,K,H,A,I,J)$ cache-aided Multi-Server-Multi-User (MSMU)  system is illustrated in Fig.~\ref{fig:system}, which consists of a file library of $N$ indenpendent files of length  $B$ (denoted by $W_1,\ldots,W_N\in\mathbb{F}_q^B$ for some prime power $q$), $H$ servers (denoted by $1,\ldots,H$), and $K$ users (denoted by $1,\ldots,K$), where each server is connected to the $K$ users via a dedicated shared-link.  \yq{
Among the $H$ servers, assume the presence of at most $A$ adversarial and colluding servers, denoted by $\mathcal{A}$, who will not collaborate in the normal operational process.} 


A central client that is reliable and safe generates the content  stored at each server $h$, denoted by $Z_h$, from the $N$ files and a random variable $U$  from some alphabet $\mathcal{U}$ with an encoding function $\chi_h: \mathbb{F}_q^{NB}\times\mathcal{U}\mapsto\mathbb{F}_q^{\lfloor TB\rfloor}$,  i.e.,
\begin{IEEEeqnarray}{c}
Z_h=\chi_h(W_{[N]}, U)\in\mathbb{F}_q^{\lfloor TB\rfloor},\quad \forall\, h\in[H],\label{eqn:chi}
\end{IEEEeqnarray}
where  $T$ is the \emph{storage size} at each server\footnote{The round down function $\lfloor\cdot\rfloor$ indicates that the cache size normalized by $B$ does not exceed $T$. }.

The system operates in two phases.  

\paragraph*{Placement Phase} 
 
Each user~$k\in[K]$ has access to the central client, and  generates some random variable $P_k$ from some finite alphabet $\mathcal{P}$ and cache some content $C_k$ as a function of $P_k$, $U$ and the file library $W_{[N]}$ using some caching function $\varphi_k:\mathcal{P}\times\mathcal{U}\times \mathbb{F}_q^{NB}\mapsto \mathbb{F}_q^{\lfloor MB\rfloor}$, i.e., 
\begin{IEEEeqnarray}{c}
C_k:= \varphi_k(P_k,U,W_{[N]})\in \mathbb{F}_q^{\lfloor MB\rfloor}, \forall\, k\in[K],\label{eqn:varchi}
\end{IEEEeqnarray}
where $M$ is the \emph{memory size} of each user. 
 \yq{Notice that the strategies of the users (i.e., the caching functions $\varphi_1,\ldots,\varphi_K$)  are known by the servers,  but the calculation of the cache contents $C_1,\ldots,C_K$ are accomplished by each user individually. }

\paragraph*{Delivery Phase} 
Each user~$k\in[K]$ generates a demand $\sd{d}_k=(d_{k,1},\ldots,d_{k,N})^\top\in\mathbb{F}_q^N$, meaning that it is interested in retrieving the linear combination of the files \yq{$W_{[N]}$ with coefficient vector $\sd{d}_k$, i.e.,
\begin{IEEEeqnarray}{c}
W_{\sd{d}_k}:=\sum_{n=1}^N d_{k,n}\cdot W_n. \notag
\end{IEEEeqnarray}}
The  random variables $\sd{d}_1,\ldots,\sd{d}_K,$ $P_1,\ldots,P_K,$ $W_1,\ldots,W_{N},U$ are independent, i.e.,
\begin{IEEEeqnarray}{rl}
&H(\sd{d}_{[K]},W_{[N]},P_{[K]},U)\notag\\
&=\sum_{k\in[K]} \! H(\sd{d}_k)
+ \sum_{n\in[N]} \! H(W_n)
+ \sum_{k\in[K]} \! H(P_k)
+ H(U).\notag
\end{IEEEeqnarray}

User~$k\in[K]$   sends a query $Q_{k,h}$ of length $\ell_{k,h}$ to server $h$ for each $h\in[H]$, where the queries $Q_{k,[H]} := (Q_{k,1},\ldots,Q_{k,H})$ are generated from some query function $\kappa_{k,h}:\mathbb{F}_q^N\times \mathcal{P}\mapsto \mathbb{F}_q^{\ell_{k,h}}$, i.e., 
\begin{IEEEeqnarray}{c}
  Q_{k,h} := \kappa_{k,h} (\sd{d}_k, P_k)\in\mathbb{F}_q^{\ell_{k,h}},\forall\, h\in[H],\label{eqn:query}
\end{IEEEeqnarray}

Upon receiving the queries from all the users, 
 each server $h\in[H]$ creates a signal $X_h\in  \mathbb{F}_q^{\lfloor R_hB\rfloor}$ as
\begin{IEEEeqnarray}{c}
X_h:=\phi_h(Z_h,Q_{[K],h}),\label{eqn:XhPhi}
\end{IEEEeqnarray}
where  $\phi_h:\mathbb{F}_q^{\lfloor TB\rfloor}\times \mathbb{F}_q^{\sum_{k\in[K]}\ell_{k,h}}\mapsto \mathbb{F}_q^{\lfloor R_hB\rfloor}$ is the the encoding function of server $h$. \yq{Then each server $h\in[H]\backslash\mathcal{A}$ sends $X_h$ to the users, while each adversarial server $h\in\mathcal{A}$ sends an arbitrary signal $\widetilde{X}_h\in\mathbb{F}_q^{\lfloor TB\rfloor}$ to the users, such that 
\begin{IEEEeqnarray}{c}
I(W_{[N]},Z_{[H]\backslash\mathcal{A}},Q_{[K],[H]\backslash\mathcal{A}},\sd{d}_{[K]}; \widetilde{X}_{\mathcal{A}}\,|\,Z_{\mathcal{A}},Q_{[K],\mathcal{\mathcal{A}}})=0.\label{XA:I}
\end{IEEEeqnarray}  
That is, the signals from the adversarial servers can be formed by all the adversarial servers, but contains no information about $(W_{[N]},Z_{[H]\backslash\mathcal{A}},Q_{[K],[H]\backslash\mathcal{A}},\sd{d}_{[K]})$ aside from their available informration $(Z_{\mathcal{A}},Q_{[K],\mathcal{A}})$.}

The quantity $R_h$ is referred to as the \emph{communication load of server $h$}. The \emph{communication load} of the system is defined as
\begin{IEEEeqnarray}{c}
R:=\max_{h\in[H]} R_h. \notag
\end{IEEEeqnarray}

  The system is designed to satisfy the following constraints:
\yq{
\begin{subequations}\label{scheme:constraint}
\begin{enumerate}  
\item \emph{Server Security:} 
Any $I$ servers should be oblivious to the files $W_{[N]}$, even if they collude, i.e.,
\begin{IEEEeqnarray}{c}
I(W_{[N]}; Z_{\mathcal{I}})=0,\quad\forall\, \mathcal{I}\subseteq[H],  |\mathcal{I}|=I.\label{eqn:server-constraint:a}
\end{IEEEeqnarray}
In particluar, the adversarial servers $\mathcal{A}$ can not jointly obtain any information about the files $W_{[N]}$, since $|\mathcal{A}|=A\leq I$.
\item \emph{Robust Recovery:} The whole files $W_{[N]}$ can be recovered from any $J$ servers, even if the adversarial servers conclude to provide erroneous information, i.e., 
\begin{IEEEeqnarray}{c}
H(W_{[N]}|\,Z_{\mathcal{J}\backslash\mathcal{A}},\widetilde{Z}_{\mathcal{J}\cap\mathcal{A}})=0,\quad
\forall\,\mathcal{A}, \mathcal{J}\subseteq[H],|\mathcal{A}|\leq A, |\mathcal{J}|=J,\IEEEeqnarraynumspace\label{eqn:server-constraint:b}
\end{IEEEeqnarray}
where $\widetilde{Z}_h\in\mathbb{F}_q^{\lfloor TB\rfloor}$ is arbitrary content provided by server $h\in\mathcal{A}$ that  satisfies
\begin{IEEEeqnarray}{c}
I(W_{[N]},Z_{[H]\backslash\mathcal{A}};\widetilde{Z}_{\mathcal{A}}\,|\,Z_{\mathcal{A}})=0.\notag
\end{IEEEeqnarray}
That is, the erroneous content $\widetilde{Z}_{\mathcal{A}}$ can be jointly generated by the adversarial servers $\mathcal{A}$, but does not contain any information about $(W_{[N]},Z_{[H]\backslash\mathcal{A}})$ aside from their content $Z_{\mathcal{A}}$.
\item \emph{Robust Decoding:} Each user $k$ can decode its demand from any $J$-subset of signals with the help of the content in its memory, even if the adversarial servers send erroneous signals,i.e.,  
\begin{IEEEeqnarray}{c}
H(W_{\sd{d}_k}\,|\,X_{\mathcal{J}\backslash\mathcal{A}},\widetilde{X}_{\mathcal{J}\cap\mathcal{A}},\sd{d}_k,C_k)=0,\quad \forall\,\mathcal{A}, \mathcal{J}\subseteq[H]:|\mathcal{A}|\leq A, |\mathcal{J}|= J.\label{eqn:correctness}
\end{IEEEeqnarray} 
\item \emph{Signal Security:} A wiretapper observing all the delivery signals  can not obtain any information about the contents of the library files
\begin{IEEEeqnarray}{c}
I(W_{[N]};X_{[H]\backslash \mathcal{A}},\widetilde{X}_\mathcal{A})=0.  \label{eqn:security}
\end{IEEEeqnarray}
\item \emph{Demand Privacy:} Any subset of users together with the servers can not jointly learn any information on the demands of the other users, regardless of the file realizations, i.e.,  
\begin{IEEEeqnarray}{c}
I(\sd{d}_{[K]\backslash\mathcal{S}};C_{\mathcal{S}},\sd{d}_{\mathcal{S}},Q_{[K],[H]},Z_{[H]}\,|\,W_{[N]})=0, \quad \forall\, \mathcal{S}\subseteq[K].\label{eqn:user:privacy}
\end{IEEEeqnarray}
\end{enumerate}
\end{subequations}

}

\begin{definition}
A \emph{Memory-Storage-Communication (MSC)}  triple $(M,T,R)\in[1,N]\times\mathbb{R}^+\times \mathbb{R}^{+}$ is said to be $B$-achievable if, for any $\epsilon>0$,  there exists a scheme satisfying all the conditions in~\eqref{eqn:server-constraint:a}, \eqref{eqn:server-constraint:b}, \eqref{eqn:correctness}--\eqref{eqn:user:privacy} with memory size less than $M+\epsilon$, storage size less than $T+\epsilon$, and communication load less than $R+\epsilon$ with file-length $B$.  The MSC region of the system is defined as
\begin{IEEEeqnarray}{c}
\mathcal{E}=\{(M,T,R):(M,T,R)~\mbox{is achievable}.\}.\notag
\end{IEEEeqnarray}
\end{definition}
 The main objective of this paper is to characterize the MSC region $\mathcal{E}$. 
For simplicity, we call a valid scheme \yq{that satsifies} the constraints in \eqref{scheme:constraint} a Robust, Secure, Private Scalar Linear Function Retrieval from Blind and  Adversarial Servers (RSP-LFR-BA) scheme. 
For a given scheme, we are also interested in its subpacketization level, which is defined as the number of packets each file has to be partitioned into in order to implement the scheme. Throughout this paper, we consider the case $N\geq 2$, since demand privacy is impossible for $N=1$ (i.e., there is only one possible file to be demanded).
\yq{\begin{remark} Notice that the files $W_{[N]}$ and the randomness $U$ are held by the central client, but not by any server or any user.  The random variable $P_k$ is generated by user $k$ who computes its cache content $C_k$, but it is kept private to all the other users,  servers, and the client. The only contents stored by server $h$ and user $k$ are $Z_h$ and $C_k$, respectively.  The wiretapper is assumed to have access to the data exchanged between the servers and the users, thus, it has access to the signals $X_{[H]\backslash\mathcal{A}},\widetilde{X}_{\mathcal{A}}$, but not to the random variables $U, P_{[K]}$ or the files $W_{[N]}$. 
 \end{remark}}

\begin{remark} Notice that, by \eqref{eqn:user:privacy} 
\begin{IEEEeqnarray}{rCl}
&&\yq{I(\sd{d}_{[K]};X_{[H]\backslash\mathcal{A}},\widetilde{X}_\mathcal{A}\,|\,W_{[N]})}\notag\\
&\leq&I(\sd{d}_{[K]};X_{[H]\backslash\mathcal{A}},\widetilde{X}_{\mathcal{A}},Q_{[K],[H]},Z_{[H]}\,|\,W_{[N]})\notag\\
&=&I(\sd{d}_{[K]};Q_{[K],[H]},Z_{[H]}\,|\,W_{[N]})+I(\sd{d}_{[K]};X_{[H]\backslash\mathcal{A}},\widetilde{X}_{\mathcal{A}}\,|\,W_{[N]},Q_{[K],[H]},Z_{[H]})\notag\\
&=&I(\sd{d}_{[K]};\widetilde{X}_{\mathcal{A}}\,|\,W_{[N]},Q_{[K],[H]},Z_{[H]})\label{eqn:ref:b}\\
&=&0,\label{eqn:ref:c}
\end{IEEEeqnarray}
where \eqref{eqn:ref:b} follows from \eqref{eqn:user:privacy} and \eqref{eqn:XhPhi}, and \eqref{eqn:ref:c} follows from \eqref{XA:I}. Together with \eqref{eqn:security},  it holds that   
\begin{IEEEeqnarray}{l}
I(W_{[N]},\sd{d}_{[K]};X_{[H]\backslash\mathcal{A}},\widetilde{X}_\mathcal{A})=0,\notag
\end{IEEEeqnarray}
that is, the wiretapper having access to the signals \yq{$X_{[H]\backslash\mathcal{A}},\widetilde{X}_\mathcal{A}$} in fact can not obtain any information on both the files and the demands of the users.
\end{remark}
\begin{remark}\label{remark:M}
It was proved in~\cite{Security} that, in order to guarantee the  conditions in~\eqref{eqn:correctness} and~\eqref{eqn:security} simultaneously, the memory size $M$ has to be no less than one.
Thus the MSC region is defined for $M\in[1,N]$.
\end{remark}
\section{PDAs and A Toy Example}\label{sec:PDA:example}
We will construct our RSP-LFR-BA scheme for any given PDA~\cite{Yan2017PDA}, which was introduced  to reduce the subpacketization in coded caching in the single server system without any security or privacy constraints.
In this section, we first review the definition of PDA, and then give an 
example to highlight the key ideas in the design of our RSP-LFR-BA scheme. 
The general construction will be given in Section \ref{sec:scheme}.
\subsection{Placement Delivery Array}
\begin{definition}[PDA~\cite{Yan2017PDA}]\label{def:PDA} For given $K,F\in\mathbb{N}^+$ and $Z,S\in\mathbb{N}$,  an $F\times K$ array
  $\mathbf{A}=[a_{j,k}]$, $j\in [F], k\in[K]$, composed of
  $Z$ specific symbols ``$*$"  in each column and some ordinary symbols $1,\ldots, S$,
  each occurring at least once,  is called a $(K,F,Z,S)$ PDA, if, for
  any two distinct entries $a_{j,k}$ and $a_{j',k'}$,   we have
  $a_{j,k}=a_{j',k'}=s$, for some ordinary symbol $s\in[S]$ only if
  \begin{enumerate}
     \item [a)] $j\ne j'$, $k\ne k'$, i.e., they lie in distinct rows and distinct columns; and
     \item [b)] $a_{j,k'}=a_{j',k}=*$, i.e., the corresponding $2\times 2$  sub-array formed by rows $j,j'$ and columns $k,k'$ must be of the following form
  \begin{IEEEeqnarray}{c}
    \left[\begin{array}{cc}
      s & *\\
      * & s
    \end{array}\right]~\textrm{or}~
    \left[\begin{array}{cc}
      * & s\\
      s & *
    \end{array}\right].\notag
  \end{IEEEeqnarray}
   \end{enumerate}
   \end{definition}
It was shown in \cite{Yan2017PDA} that, with a given $(K,F,Z,S)$ PDA,  there exists an associated coded caching scheme in the single server system without any security or privacy constraint.  The parameter $K$ is the number of users, $F$ is the number of packets each file is split into (i.e., subpacketization), $Z$ is the number of uncoded packets from each file stored at each user, and $S$ is the number of coded multicast signals.  In our model, each file will first split  into $L=J-I-2A$ equal-size subfiles, and those implications will be used on the subfiles. 
\subsection{A Toy RSP-LFR-BA Example from PDAs}\label{sec:example}

We derive here a RSP-LFR-BA scheme associated to the $(K,F,Z,S)=(3,3,1,3)$ PDA 
\begin{IEEEeqnarray}{c}
\mathbf{A}=\left[\begin{array}{ccc}
        * & 1 &2\\
        1 & * &3\\
        2& 3&*
      \end{array}
\right]
\label{eqn:pippo}
\end{IEEEeqnarray}
for a $(N,K,H,A,I,J)=(4,3,6,1,1,5)$ MSMU system. 

Let the four original files be $W_1,W_2,W_3,W_4\in\mathbb{F}_q^B$. Firstly, each file $W_n, n\in[4]$ is split into
\yq{ 
\begin{IEEEeqnarray}{c}
L\triangleq J-I-2A=2
\end{IEEEeqnarray}
 }equal-size subfiles,
\begin{IEEEeqnarray}{c}
 W_n=(W_{n,1},W_{n,2}),\notag
\end{IEEEeqnarray}
and each subfile $W_{n,l}, l\in[2]$ is further split into $F=3$ equal-size packets:
\begin{IEEEeqnarray}{c}
W_{n,l}=(W_{n,l,1},W_{n,l,2},W_{n,l,3}).\label{eqn:Wnl:exam}
\end{IEEEeqnarray}
\yq{The system generates two group of random variables. The first group contains $N=4$ uniform and independent random variables 
$\{\Delta_n:n\in[4]\}$  from $\mathbb{F}_q^{\frac{B}{2}}$  (the range of a subfile), that will be used as noises to encode the subfiles from the $N=4$ files separately. The second group contains $(L+I)S=9$ uniform and independent random variables $\{V_{l,s}:l\in[2],s\in[3]\}\yq{\cup\{\Lambda_s:s\in[3]\}}$ from $\mathbb{F}_q^{\frac{B}{6}}$ (the range of a packet),  where $V_{l,s}$ are used as keys to protect the coded signals, but $\Lambda_s$ are used as noises to protect the keys. To be concret, two groups of polynomials are constructed: }
\begin{subequations}\label{eqn:poly:exam}
\begin{IEEEeqnarray}{rCl}
\widetilde{W}_n(x)&=&W_{n,1}+W_{n,2}x+\Delta_nx^2,\quad \forall\, n\in[4],\label{eqn:secret-sharing:exam}\\
\widetilde{V}_s(x)&=&V_{1,s}+V_{2,s}x\yq{+\Lambda_sx^2},\quad\forall\, s\in[3].\label{eqn:secret-sharing:exam:b}
\end{IEEEeqnarray}
\end{subequations}
Notice that, in accordance with \eqref{eqn:Wnl:exam}, for each $n\in[3]$,  $\widetilde{W}_n(x)$ and $\Delta_n$ can be decomposed into $3$ equal-size components, i.e.,  $\widetilde{W}_n(x)=(\widetilde{W}_{n,1}(x),\widetilde{W}_{n,2}(x),\widetilde{W}_{n,3}(x))$ and $\Delta_n=(\Delta_{n,1},\Delta_{n,2},\Delta_{n,3})$, where
\begin{IEEEeqnarray}{c}
\widetilde{W}_{n,j}(x)=W_{n,1,j}+W_{n,2,j}x+\Delta_{n,j}x^2,\quad j\in[3].\notag
\end{IEEEeqnarray}
Let $\alpha_1,\alpha_2,\alpha_3,\alpha_4,\alpha_5,\alpha_6$ be $H=6$ distinct non-zero elements in $\mathbb{F}_q$. The system operates as follows:
\paragraph*{Server Storage Design} Each server $h\in[6]$ stores the evaluations of the polynomials in \eqref{eqn:poly:exam}:
\begin{IEEEeqnarray}{c}
Z_h=\{\widetilde{W}_n(\alpha_h):n\in[4]\}\cup\{\widetilde{V}_s(\alpha_h):s\in[3]\}. \notag
\end{IEEEeqnarray} 
\paragraph*{Placement Phase} Each user~$k\in[3]$ generates a random vector $\sd{p}_k=(p_{k,1},p_{k,2},p_{k,3},p_{k,4})^\top\in\mathbb{F}_q^4$.  The cache content of the user~$k$ is composed of $\sd{p}_k$ and the (un)coded packets in the corresponding column in Table \ref{table:Z}.
\begin{table}[htbp]
\small
\centering
\scalebox{1}{\begin{threeparttable}
\caption{The cached contents of users$^{\dagger}$ according to $\mathbf{A}$ in~\eqref{eqn:pippo}.}\label{table:Z}
\begin{tabular}{cccc}
  \bottomrule
 Row &User~$1$& User~$2$& User~$3$\\\hline
$1$ &$W_{[4],[2],1}$&$W_{\sd{p}_2,[2],1}\oplus V_{[2],1}$&$W_{\sd{p}_3,[2],1}\oplus V_{[2],2}$\\
 $2$&$W_{\sd{p}_1,[2],2}\oplus V_{[2],1}$& $W_{[4],[2],2}$& $W_{\sd{p}_3,[2],2}\oplus V_{[2],3}$\\ $3$&$W_{\sd{p}_1,[2],3}\oplus V_{[2],2}$ &$W_{\sd{p}_2,[2],3}\oplus V_{[2],3}$ &$W_{[4],[2],3}$\\
 \bottomrule
\end{tabular}
\begin{tablenotes}
\item[$\dagger$]
\footnotesize{In addition, each user~$k\in[3]$ caches $\sd{p}_k$.}
\end{tablenotes}
\end{threeparttable}}
\end{table}
The packets $W_{[4],[2],j}$ are associated to the $j$-th row of $\mathbf{A}$ in~\eqref{eqn:pippo} and user~$k$ is associated to the $k$-th column of $\mathbf{A}$. The packets in the $j$-th row of Table \ref{table:Z} of  user~$k$ are created according to the entry $a_{j,k}$ of $\mathbf{A}$ in~\eqref{eqn:pippo}\footnote{As we assume that the users have access to all the servers in the placement phase, each user can fill its cache with arbitrary function of the files $W_{[4]}$ and the random variables $\Delta_{[4]},\Lambda_{[3]},V_{[2],[3]}$. } : if $a_{j,k}=*$, user~$k$ caches $NL=8$ uncoded packets $W_{[4],[2],j}$, otherwise it caches $L=2$ coded packets $W_{\sd{p}_{k},[2],j}\oplus V_{[2],a_{j,k}}$. 

\paragraph*{Delivery Phase}
Assume that user~$1,2,3$ demands the linear combinations 
$W_{\sd{d}_1},W_{\sd{d}_2}$ and $W_{\sd{d}_3}$, respectively, where $\sd{d}_1,\sd{d}_2,\sd{d}_3\in\mathbb{F}_q^4$. Each user~$k\in[3]$ sends $\sd{q}_k=\sd{p}_k\oplus \sd{d}_k$ to all the servers as queries. Upon receiving the query vectors $\sd{q}_{[3]}$,  each server $h\in[6]$ creates a signal $X_h$, where $X_h$ is composed of the query vectors $\sd{q}_{[3]}$ and $S=3$ coded packets  associated to the ordinary symbols $s=1,2,3$ respectively:
\begin{IEEEeqnarray}{l}
s=a_{1,2}=a_{2,1}=1:
\quad \widetilde{Y}_1(\alpha_h)=\widetilde{V}_1(\alpha_h)+\widetilde{W}_{\sd{q}_1,2}(\alpha_h)+\widetilde{W}_{\sd{q}_2,1}(\alpha_h),\notag\\
s=a_{1,3}=a_{3,1}=2:
\quad \widetilde{Y}_2(\alpha_h)=\widetilde{V}_2(\alpha_h)+\widetilde{W}_{\sd{q}_1,3}(\alpha_h)+\widetilde{W}_{\sd{q}_3,1}(\alpha_h),\notag\\
s=a_{2,3}=a_{3,2}=3:
\quad \widetilde{Y}_3(\alpha_h)=\widetilde{V}_3(\alpha_h)+\widetilde{W}_{\sd{q}_2,3}(\alpha_h)+\widetilde{W}_{\sd{q}_3,2}(\alpha_h),\notag
\end{IEEEeqnarray}
where a polynomial $\widetilde{W}_{\sd{b},j}(x)$  is defined for each $\sd{b}=(b_1,b_2,b_3,b_4)\in\mathbb{F}_q^4$ and $j\in[3]$ as:
\begin{IEEEeqnarray}{c}
\widetilde{W}_{\sd{b},j}(x)\triangleq b_{1}\widetilde{W}_{1,j}(x)+b_2\widetilde{W}_{2,j}(x)+b_3\widetilde{W}_{3,j}(x)+b_4\widetilde{W}_{4,j}(x).\notag
\end{IEEEeqnarray}
\yq{Among the servers, any normal server would send $X_h$ to the users, while the adversarial server, denoted by $\tilde{h}\in[6]$, creates an arbitrary signal $\widetilde{X}_{\tilde{h}}$ of the same size as $X_h$, based on its own cache content $Z_{\tilde{h}}$, and send it to the users.  }

\paragraph*{Robust Decoding} Let's take $s=1$ to show how the users utilize the signals to decode. Consider the polynomial
\begin{IEEEeqnarray}{rCl}
&&\widetilde{Y}_1(x)\notag\\
&=&\widetilde{V}_1(x)+\widetilde{W}_{\sd{q}_1,2}(x)+\widetilde{W}_{\sd{q}_2,1}(x)\notag\\
&=&\widetilde{V}_1(x)+\sum_{n=1}^4q_{1,n}\widetilde{W}_{n,2}(x)+\sum_{n=1}^4q_{2,n}\widetilde{W}_{n,1}(x)\notag\\
&=&(V_{1,1}+W_{\sd{q}_1,1,2}+W_{\sd{q}_2,1,1})+(V_{2,1}+W_{\sd{q}_1,2,2}+W_{\sd{q}_{2},2,1})x+(\Lambda_1+\Delta_{\sd{q}_1,2}+\Delta_{\sd{q}_2,1})x^2\label{eqn:Y1:x}\\
&=&Y_{1,1}+Y_{1,2}x+\widetilde{\Delta}_1x^2,\label{eqn:Y1:poly}
\end{IEEEeqnarray}
\yq{where 
\begin{IEEEeqnarray}{rCl}
Y_{1,1}&\triangleq&V_{1,1}+W_{\sd{q}_1,1,2}+W_{\sd{q}_2,1,1},\\
Y_{1,2}&\triangleq&V_{2,1}+W_{\sd{q}_1,2,2}+W_{\sd{q}_{2},2,1},\\
\widetilde{\Delta}_1&\triangleq&\Lambda_1+\Delta_{\sd{q}_1,2}+\Delta_{\sd{q}_2,1}.
\end{IEEEeqnarray}}
Notice that, the signal $\widetilde{Y}_1(\alpha_h)$ sent by server $h$ is the evaluation of $\widetilde{Y}_1(x)$ at $\alpha_h$. \yq{Moreover, by \eqref{eqn:Y1:poly}, the signals $\{\widetilde{Y}_1(\alpha_h)\}_{h\in[6]}$ is a $3$ dimensional  Maximum Distance Separable (MDS) codeword of length $6$:
\begin{IEEEeqnarray}{c}
\big(\widetilde{Y}_1(\alpha_1),\widetilde{Y}_1(\alpha_2),\widetilde{Y}_1(\alpha_3),\widetilde{Y}_1(\alpha_4),\widetilde{Y}_1(\alpha_5),\widetilde{Y}_1(\alpha_6)\big)=(Y_{1,1},Y_{1,2},\widetilde{\Delta}_1)\left[\begin{array}{cccccc}
1&1&1&1&1&1\\
\alpha_1&\alpha_2&\alpha_3&\alpha_4&\alpha_5&\alpha_6\\
\alpha_1^2&\alpha_2^2&\alpha_3^2&\alpha_4^2&\alpha_5^2&\alpha_6^2
\end{array}\right].\IEEEeqnarraynumspace
\end{IEEEeqnarray}
By the capability of MDS code,  each user can decode $(Y_{1,1},Y_{1,2},\widetilde{\Delta}_1)$ if it receives $J=5$ signals that contains $A=1$ erroneous signal. This indicates that with signals from arbitrary $5$ servers,  the user can the signals as listed in Table \ref{table:S}, even if there exists an adversarial server that sends erroneous signals to it.} 
\begin{table*}[htbp]
\centering
\scalebox{1}{
\begin{threeparttable}
\caption{The signals a user can decode from the signals of any $5$ servers$^\dagger$ according to $\mathbf{A}$.}\label{table:S}
\begin{tabular}{cccc}
\toprule
$s$&Subfiles $1$&Subfiles $2$& Noises\\\hline
$1$&$V_{1,1}+ W_{\sd{q}_1,1,2}+W_{\sd{q}_2,1,1}$&$V_{2,1}+ W_{\sd{q}_1,2,2}+W_{\sd{q}_2,2,1}$&$
\Lambda_1+\Delta_{\sd{q}_1,2}+\Delta_{\sd{q}_2,1}$\\
 $2$&$V_{1,2}+ W_{\sd{q}_1,1,3}+ W_{\sd{q}_3,1,1}$&$V_{2,2}+ W_{\sd{q}_1,2,3}+ W_{\sd{q}_3,2,1}$&$\Lambda_2+\Delta_{\sd{q}_1,3}+\Delta_{\sd{q}_3,1}$\\
 $3$&$V_{1,3}+ W_{\sd{q}_2,1,3}+ W_{\sd{q}_3,1,2}$&$V_{2,3}+ W_{\sd{q}_2,2,3}+W_{\sd{q}_3,2,2}$&$\Lambda_3+\Delta_{\sd{q}_2,3}+\Delta_{\sd{q}_3,2}$\\
 \bottomrule
\end{tabular}
\begin{tablenotes}
\item[$\dagger$]
\footnotesize{In addition, each server $h\in[6]$ transmits the query vectors $\sd{q}_{[3]}$.}
\end{tablenotes}
\end{threeparttable}}
\end{table*}

Upon obtaining the signals in Table \ref{table:S}, each user~$k\in[3]$ can proceed with the decoding process for each subfile $l\in[2]$ as in~\cite{Y:D:Privacy}. For example, for subfile 1,  as $a_{1,2}=a_{2,1}=1$, 
user~$1$ can decode $W_{\sd{d}_1,1,2}$ and user~$2$ can decode $W_{\sd{d}_2,1,1}$
from the signal $V_{1,1}+ W_{\sd{q}_1,1,2}+ W_{\sd{q}_2,1,1}$, since
\begin{itemize}
\item user $1$ has cached the superposition key $W_{\sd{p}_1,1,2}+V_{1,1}$ and can compute $W_{\sd{q}_2,1,1}$ from its cached subfiles $W_{[4],1,1}$ and the query vector $\sd{q}_2$, and thus
\begin{IEEEeqnarray}{l}
(V_{1,1}+ W_{\sd{q}_1,1,2}+ W_{\sd{q}_2,1,1})-(W_{\sd{p}_1,1,2}+V_{1,1})-W_{\sd{q}_2,1,1}\notag\\
=W_{\sd{q}_1,1,2}-W_{\sd{p}_1,1,2}\notag\\
=W_{\sd{d}_1,1,2},\notag
\end{IEEEeqnarray}
where the last step follows from $\sd{q}_1=\sd{p}_1+\sd{d}_1$;
\item user $2$ has cached the superposition key $W_{\sd{p}_2,1,1}+V_{1,1}$ and can compute $W_{\sd{q}_1,1,2}$ from its cached subfiles $W_{[4],1,2}$ and the query vector $\sd{q}_1$, and thus
\begin{IEEEeqnarray}{l}
(V_{1,1}+ W_{\sd{q}_1,1,2}+ W_{\sd{q}_2,1,1})-(W_{\sd{p}_2,1,1}+V_{1,1})-W_{\sd{q}_1,1,2}\notag\\
=W_{\sd{q}_2,1,1}-W_{\sd{p}_2,1,1}\notag\\
=W_{\sd{d}_2,1,1},\notag
\end{IEEEeqnarray}
where the last step follows from $\sd{q}_2=\sd{p}_2+\sd{d}_2$.
\end{itemize}
One can verify that each user~$k\in[3]$ can decode all the remaining packets $W_{\sd{d}_k,[2],[3]\backslash\{k\}}$ from its stored contents,  the signals in Table \ref{table:S} and the query vectors $\sd{q}_{[3]}$.      

\paragraph*{Security and Privacy Conditions} In addition to the robust decoding, the other conditions in \eqref{scheme:constraint} are also guaranteed: In fact, the secret sharing techique \cite{Shamir1979} to encode files with polynomials in \eqref{eqn:secret-sharing:exam} guarantees the $I$-Security,  $J$-Robust Recovery; Signal Security is guaranteed by the $I$-Security and the fact that each signal is accompanied by a uniformly distributed key, and Demand Privacy is guaranteed by the fact that the queries are formed by the demands accompanied with uniformly distributed vectors.

\paragraph*{Performance} Recall that each packet is of length $\frac{B}{6}$. 
Each user caches $12$ packets and $1$ vector in $\mathbb{F}_q^4$, whose length does not scale with $B$. Thus the needed memory is $M=12\times\frac{1}{6}=2$ files.  Each server stores  $N=4$ subfiles, each of length $\frac{B}{2}$, and $3$ coded packets, each of length $\frac{B}{6}$, thus the storage size is $T=4\times\frac{1}{2}+3\times \frac{1}{6}=\frac{5}{2}$. 
Each server transmits $3$ packets and $3$ vectors in $\mathbb{F}_q^4$, thus the achieved load is $R= 3\times\frac{1}{6}=\frac{1}{2}$ files. Hence, the scheme achieves the memory-storage-load triple  $(M,T,R)=\big(2,\frac{5}{2},\frac{1}{2}\big)$. Since each file is split into $6$ equal-size packets, the subpacketization level  is $LF=6$.

\begin{remark} The key idea to merge the superposition key with the secret sharing is the design of the polynomials in \eqref{eqn:poly:exam}, where the polynimials in \eqref{eqn:secret-sharing:exam:b} encodes the secret keys into polynomials such that the secret keys are aligned with the signals in the Shamir's secret sharing polynomials in \eqref{eqn:secret-sharing:exam}. In this way, the security keys can be used to protect the multicast signals, even in a coded form, see \eqref{eqn:Y1:x}. 
\end{remark}

\section{Main Results}\label{sec:main}
\subsection{PDA based RSP-LFR-BA Schemes}
With any given PDA, we will construct an associated RSP-LFR-BA scheme. We will prove the following theorem by presenting and analyzing the construction in Section \ref{sec:scheme}. The scheme is a direct generalization of the illustrative example in Section \ref{sec:example}. \yq{For ease of depiction, we define 
\begin{IEEEeqnarray}{c}
L:=J-I-2A. \label{def:L}
\end{IEEEeqnarray}
We will see that $L$ is the number of subfiles that each file is split into in the proposed scheme. 
}
\begin{theorem}\label{thm:PDA} For a given $(K,F,Z,S)$ PDA $\mathbf{A}$, there exists an associated RSP-LFR-BA scheme that achieves the  MSC triple
\begin{IEEEeqnarray}{c}
\big(M_{\mathbf{A}},T_{\mathbf{A}},R_{\mathbf{A}}\big)=\bigg(1+\frac{Z}{F}(N-1) , \frac{1}{L}\Big(N+\frac{S}{F}\Big), \frac{S}{LF} \bigg).\notag
\end{IEEEeqnarray}
with subpacketization $LF$.
\end{theorem}


The advantage of establishing a PDA framework is that, all the existing PDAs can be used to derive RSP-LFR-BA schemes. As the subpacketization of the PDA based RSP-LFR-BA scheme is $LF$, which is proportional to $F$,  PDAs with small number of rows  result low subpacketization schemes. 

Notice that, the storage size
\begin{IEEEeqnarray}{c}
T_{\bf{A}}=\frac{N}{L}+ R_{\bf A}.\label{Linear:TR}
\end{IEEEeqnarray}
In fact, from scheme description in Section \ref{sec:scheme}, $\frac{N}{L}$ is the storage size used to store the encoded version of the files, while $R(M)$ is the size of the encoded version of security keys used to protect the signal from the wiretapper, which is the same with the signal size transmitted by each server.

\yq{
\subsection{Lower Bound for RSP-LFR-BA Schemes}
For any $M\in[1,N]$, define
\begin{IEEEeqnarray}{rCl}
T^*(M)&=&\inf\left\{T:\exists\,R\in\mathbb{R}^+,,\, \mbox{s.t.}\, (M,T,R)\in\mathcal{E}\right\},\\
R^*(M)&=&\inf\left\{R:\exists\,T\in\mathbb{R}^+,\,\mbox{s.t.}\, (M,T,R)\in\mathcal{E}\right\}.
\end{IEEEeqnarray}
Theorem \ref{thm:TRbounds} characterizes a bound for $T^*(M)$ and $R^*(M)$ respectively, where the proof is deferred to Section \ref{sec:bound}. 
\begin{theorem}\label{thm:TRbounds} For any given $M\in[1,N]$, the optimal storage size $T^*(M)$ and optimal communication load satisfy
\begin{subequations}
\begin{IEEEeqnarray}{rCl}
T^*(M)&\geq&\frac{N}{L+2A},\label{T:low}\\
R^*(M)&\geq &\max_{u\in[\min\{\lfloor \frac{N}{2}\rfloor,K\}]}  \frac{u(N-u+1-uM)}{(L+2A)N}.\label{lowerboud:R}
\end{IEEEeqnarray}
\end{subequations} 
\end{theorem}
Among all PDA constructions, the MAN-PDAs that describe the Maddah-Ali and Niesen's coded caching scheme \cite{Maddah-Ali2014fundamental} is of very important. The above bounds will be used to analyze the optimality of MAN-PDA based schemes in the following. 
}
\subsection{Optimality of MAN-PDA based RSP-LFR-BA Schemes}\label{sec:MAN:main}

For any integer  $t\in[0:K]$, define the set
\begin{IEEEeqnarray}{c}
\mathbf{\Omega}_t\triangleq\{\mathcal{T}\subseteq[K]:|\mathcal{T}|=t\}.\notag
\end{IEEEeqnarray}
It was proved in~\cite{Yan2017PDA} that the following construction is a $(K,{K\choose t},{K-1\choose t-1},{K\choose t+1})$ PDA, which describes the MAN scheme in~\cite{Maddah-Ali2014fundamental}  and will be referred to as MAN-PDA in the following.

\begin{definition}[MAN-PDA]\label{def:MNPDA}
Fix any integer $t\in[0:K]$, denote the set  $\mathbf{\Omega}_t=\{\mathcal{T}_j : j\in[ {K \choose t}] \}$. Also, choose an arbitrary bijective function $\kappa_{t+1}$ from $\mathbf{\Omega}_{t+1}$
to the set $\big[{K \choose {t+1}}\big]$.
Then, define the array $\mathbf{A}_t=[a_{j,k}]$ as
	\begin{IEEEeqnarray}{c}
		a_{j,k}\triangleq \left\{\begin{array}{ll}
			*, &\textnormal{if}~k\in\mathcal{T}_j \\
			\kappa_{t+1}(\{k\} \cup \mathcal{T}_{j}), &\textnormal{if}~k\notin\mathcal{T}_j
		\end{array}
		\right..\notag
	\end{IEEEeqnarray}
\end{definition}

\begin{example}[A MAN-PDA]
Consider $K=4$, $t=2$, let $\mathcal{T}_1=\{1,2\},\mathcal{T}_2=\{1,3\},\mathcal{T}_3=\{1,4\},\mathcal{T}_4=\{2,3\},\mathcal{T}_5=\{2,4\}$ and $\mathcal{T}_6=\{3,4\}$. Let $\kappa_3$ be the lexicographic order of a subset of size  $3$ in  $\mathbf{\Omega}_3$, e.g., $\kappa_3(\{1,2,3\})=1,\kappa_3(\{1,2,4\})=2$ and $\kappa_3(\{1,3,4\})=3$ and $\kappa_3(\{2,3,4\})=4$.
The corresponding $(4,6,3,4)$ PDA is given by 
\begin{IEEEeqnarray}{c}
\mathbf{A}_2=\left[\begin{array}{cccc}
*&*&1&2\\
*&1&*&3\\
*&2&3&*\\
1&*&*&4\\
2&*&4&*\\
3&4&*&*
\end{array}
\right].\notag
\end{IEEEeqnarray}
\end{example}

The following theorem summarizes the performance of MAN-PDA and its optimality. 
\begin{theorem}\label{thm:MAN}
Let $(M,T(M),R(M))$ be the 
curve formed  by connecting the points
\begin{IEEEeqnarray}{l}
(M_t,T_t, R_t)= \notag\\
\left(1+\frac{t(N-1)}{K},\frac{1}{L}\Big(N+\frac{K-t}{t+1}\Big), \frac{K-t}{L(t+1)}\right),\quad t\in[0:K]\notag
\end{IEEEeqnarray}
sequentially, then the region 
\begin{IEEEeqnarray}{c}
\{(M,T,R):T\geq T(M), R\geq R(M)\}\label{eqn:region}
\end{IEEEeqnarray}
is achievable, where the point $(M_t,T_t,R_t)$ can be achieved with subpacketization  $L{K\choose t}$. Moreover, \yq{the achievable storage size $T(M)$ and communication load $R(M)$ satisfy}
  \yq{\begin{IEEEeqnarray}{c}
  \frac{T(M)}{T^*(M)}\leq 2\Big(1+\frac{2A}{L}\Big),\quad\frac{R(M)}{R^*(M)}\leq12\Big(1+\frac{2A}{L}\Big),\label{thm:bound}
  \end{IEEEeqnarray}}
if $K\leq N$ or $K> N, M\geq 2$.
\end{theorem}

\begin{figure}[t]
  \centering
  \includegraphics[width=0.7\columnwidth]{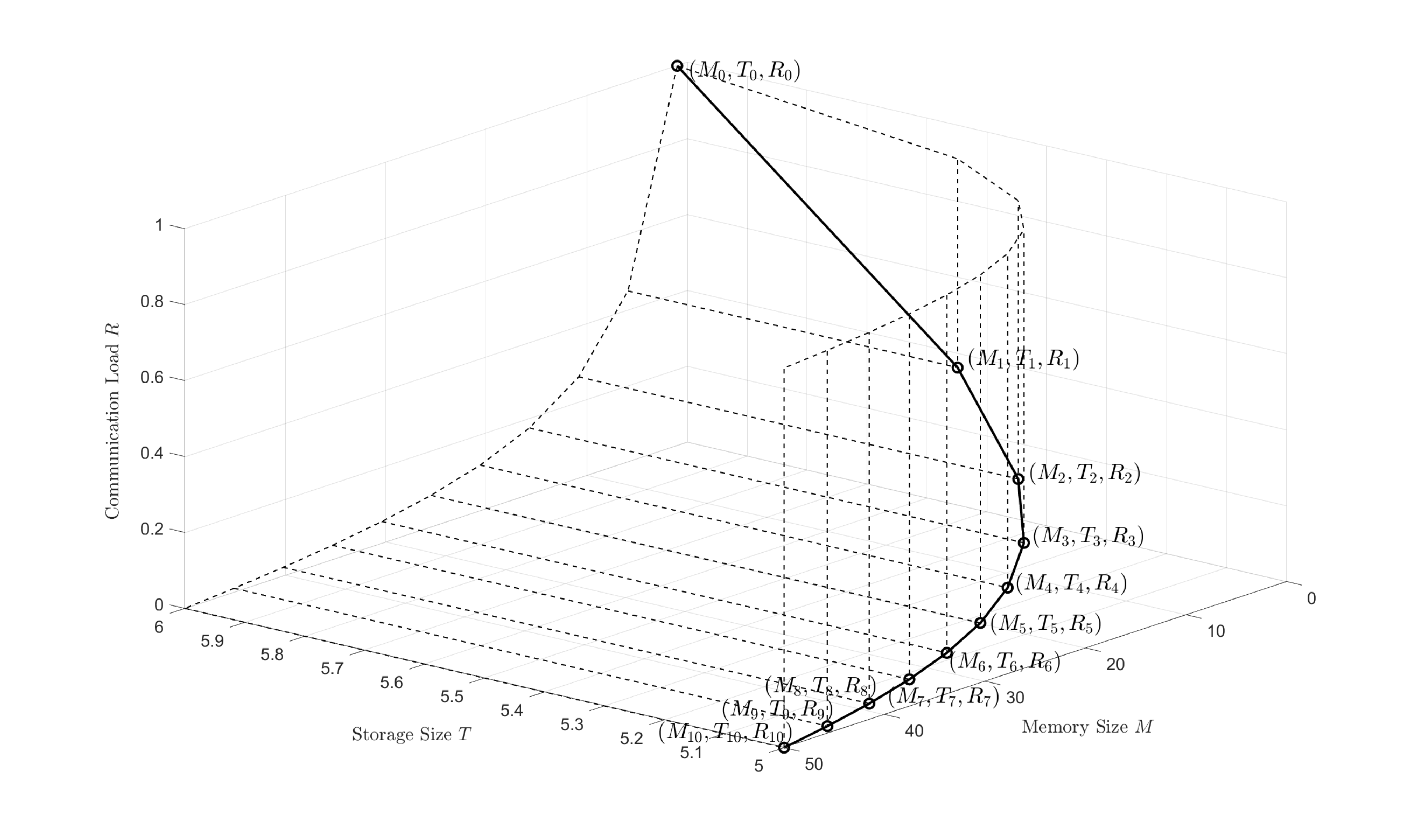}
  \caption{Achievable MSC Region of a $(50,10,20,3,10,16)$ MUMS system}\label{fig:MTR}
\vspace{-10pt}
\end{figure}

As illustrated in Fig. \ref{fig:MTR}, the curve $(M,T(M),R(M))$ is composed of $K$ line segments. Then the boundary is formed by the above surfaces starting from the curve and parallel to the $T$- and $R$-axes. Obviously, the curve  $(M,T(M),R(M)), M\in[1,N]$ forms the all Pareto-optimal points of the set, that is, we only need to verify the achievability of the curve $(M,T(M),R(M))$. In fact, the achievability of the point $(M_t,T_t,R_t)$ directly follows from Theorem~\ref{thm:PDA} and the fact that $\mathbf{A}_t$ in Definition~\ref{def:MNPDA} is a $(K,{K\choose t},{K-1\choose t-1},{K\choose t+1})$ MAN-PDA  \cite{Yan2017PDA}. 
Moreover, for general $M\in[1,N]$, $M$ lies in $[M_{t-1},M_t]$ for some $t\in[K]$, the point $(M,T(M),R(M))$ can be achieved by memory-sharing~\cite{Maddah-Ali2014fundamental} between the points $(M_{t-1},T_{t-1},R_{t-1})$ and $(M_t,T_t,R_t)$. Thus,  the achievability of the region \eqref{eqn:region} follows directly Theorem~\ref{thm:PDA} and the construction of MAN-PDA. The optimality in proved by deriving a lower bound for the $T(M)$ and $R(M)$ respectively, where the details are presented in Section~\ref{sec:MAN:PDA}.

\begin{remark}[On Unbounded Regime]\label{remark:unbound} In the regime $K> N, 1\leq M< 2$, the gap is unbounded. The main problem in small memory regime for the single server model when $K> N$ is that, if security keys are used \cite{Y:D:SP-LFR, Security}, for the point $M=1$ the best know achievable communication load is $K$, while the best known converse is $N$. Thus, it seems that
the larger  communication load when $K > N$ is mainly caused by the security
condition; closing the gap in small memory regime when $K>N$ is
an open problem in the single server system with  signal security constraint  constraint setup \cite{Security},  \cite{Y:D:SP-LFR}, and scalar linear function retrieval from MDS coded servers with signal security  constraint setups \cite{Yan2022JSAC}.
\end{remark}

\yq{\subsection{Numerical Results}\label{sec:numerical}
In this subsection, we present some numerical results.  Due to the relationship \eqref{Linear:TR}, and the fact that the bound for $\frac{T(M)}{T^*(M)}$ is tighter than that for $\frac{R(M)}{R^*(M)}$ in Theorem \eqref{thm:MAN}, we focus on the comparison of $R(M)$ and its lower bound of $R^*(M)$ derived in Theorem \ref{thm:TRbounds}.    

%
In Fig. \ref{fig:2}, we plot  $R(M)$ and the lower bound  \eqref{lowerboud:R}  for a system with $H=20,A=2,I=3,J=17$ (thus $L=J-I-2A=10$) to illustrate two regimes $K> N$ (Fig. \ref{fig2:a}, where we choose $(N,K)=(10,100)$) and $K\leq N$ (Fig. \ref{fig2:b} where we choose $(N,K)=(100,10)$).
\begin{figure}[t] \centering
\subfigure[$R(M)$ for $N=10,K=100$] {
 \label{fig2:a}
\includegraphics[width=0.48\columnwidth]{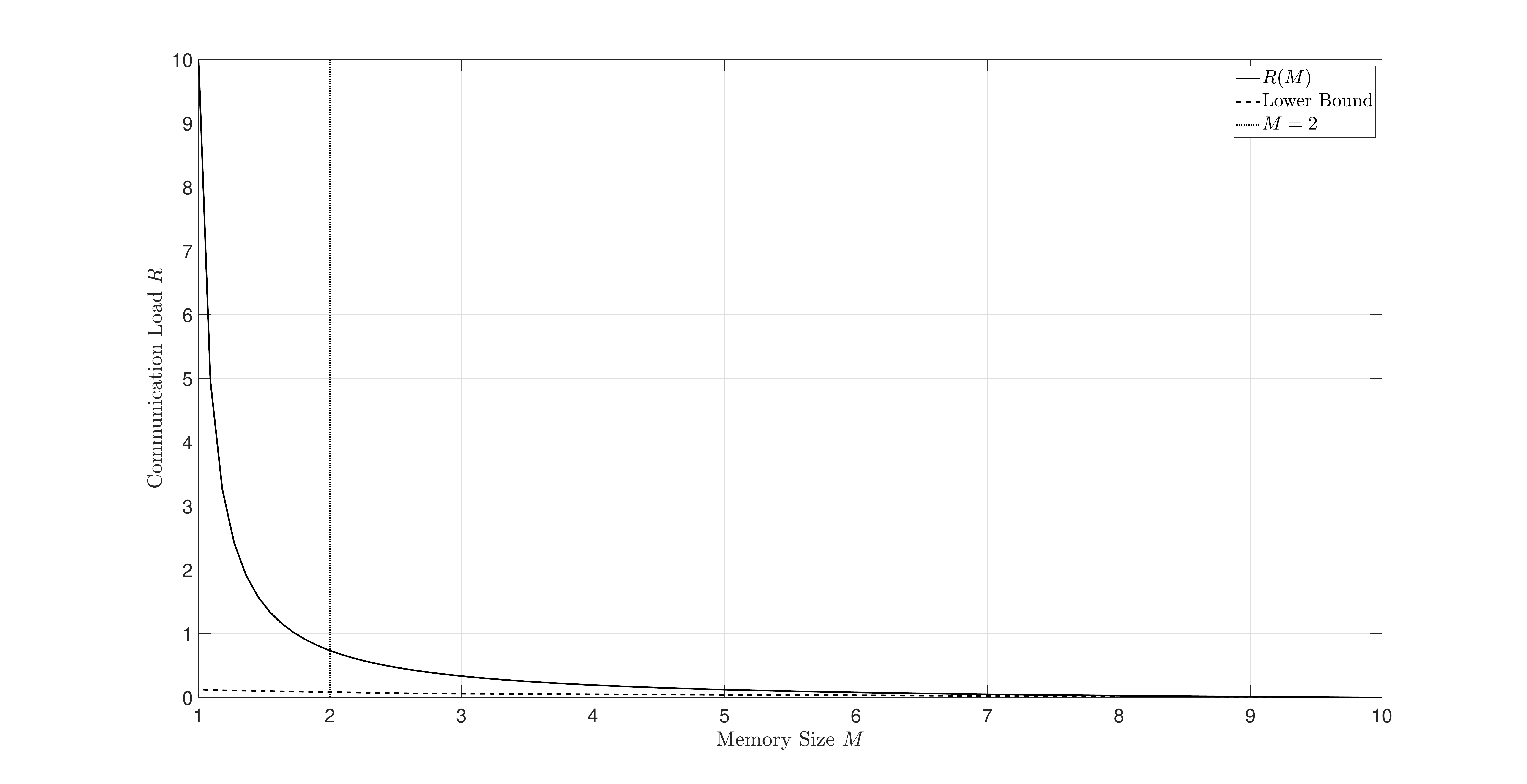}
}
\subfigure[$R(M)$ for $N=100,K=10$] {
 \label{fig2:b}
\includegraphics[width=0.48\columnwidth]{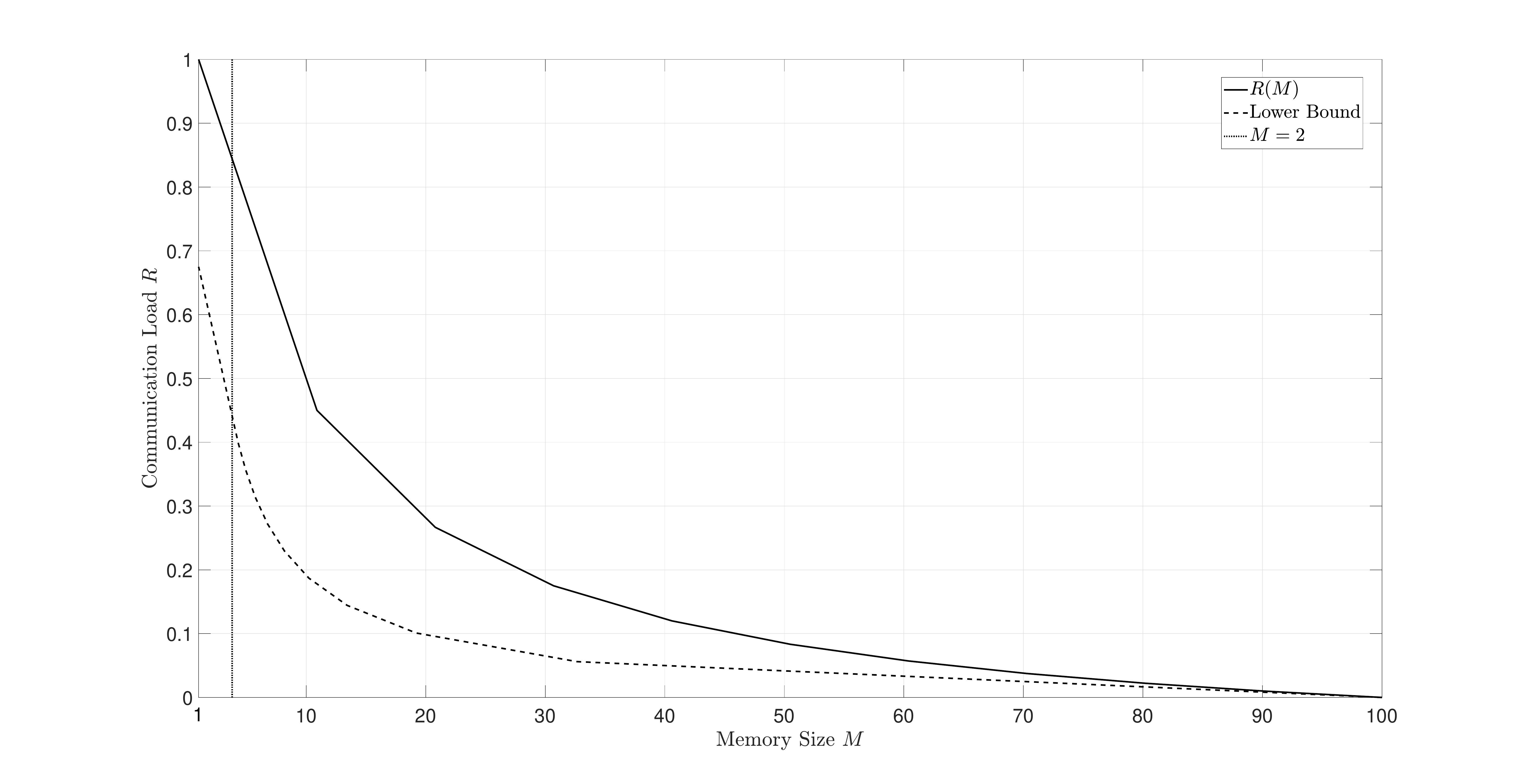}}
\caption{Communication load $R(M)$  for a MUMS system with $H=20,A=2,I=3,J=17$. }
\label{fig:2}
\end{figure}  
For comparison,  the corresponding  bound  in   \eqref{lowerboud:R} is also plotted. 
 From the figures, we have the following observations:
\begin{enumerate}
\item For the regime $K>N$, the communication load $R(M)$ scales with $K/L$ at $M=1$, and decreases dramatically around $M=1$ due to the coded multi-cast opportunities provided by coded caching. In particular, as stated  in Theorem \ref{thm:MAN}, when $M\geq 2$, $R(M)$ is close to its lower bound. 
\item For the regime $K\leq N$, 
the communication  $R(M)$ is close to its lower bound over all $M\in[1,N]$. In fact, for the chosen parameters, \eqref{eqn:2ratio} indicates that the bound can be tighted to $2.8$ for $1\leq M\leq \frac{N-2K}{2K+1}$. 
\end{enumerate}
In summary, the numerical results are consistent with Remark \ref{remark:unbound}:  the only un-bounded regime is $K>N$ and $1\leq M< 2$.}

\section{PDA Based RSPB-LFR-BA Scheme} 
\label{sec:scheme}
In this section, we prove Theorem \ref{thm:PDA} by deriving a RSP-LFR-BA scheme  from any given $(K,F,Z,S)$ PDA $\mathbf{A}=[a_{j,k}]_{F\times K}$. 
\subsection{PDA Based Scheme}
\yq{Let $L$ be the number defined in \eqref{def:L}.
}
Based on $\mathbf{A}$, the system partitions each file $W_n$  into $LF$ equal-size packets as follows:
\begin{itemize}
\item Each file $W_n$ is first patitioned into $L$ subfiles
\begin{IEEEeqnarray}{c}
W_n=(W_{n,1},\ldots,W_{n,L}),\quad\forall\, n\in[N],\label{eqn:partiton:subfile}
\end{IEEEeqnarray}
where $W_{n,l}\in\mathbb{F}_q^{\frac{B}{L}}$ is the $l$-th subfile of $W_n$ for all $l\in[L]$. 
\item Each subfile $W_{n,l}$ is further partitioned into $F$ packets
\begin{IEEEeqnarray}{c}
W_{n,l}=(W_{n,l,1},\ldots,W_{n,l,F}),~~\forall\, n\in[N],l\in[L],\IEEEeqnarraynumspace\label{eqn:partition:packets}
\end{IEEEeqnarray}
where $W_{n,l,j}\in\mathbb{F}_q^{\frac{B}{LF}}$ is the $j$-th packet of the subfile $W_{n,l}$ for all $j\in[F]$. 
\end{itemize}

The system generates $NI$ random subfiles, denoted by \yq{$\Delta_{1,1},\ldots,\Delta_{N,I}$}, and \yq{$(L+I)S$} random packets, denoted by $V_{1,1},V_{1,2},\ldots,V_{L,S},\yq{\Lambda_{1,1},\Lambda_{1,2},\ldots,\Lambda_{I,S}}$, where the random subfiles $\Delta_{n,i}$ are uniformally distributed over $\mathbb{F}_q^{\frac{B}{L}}$, and the random packets $V_{l,s}$ and $\Lambda_{i,s}$ are uniformally distributed over $\mathbb{F}_q^{\frac{B}{LF}}$. That is, the random variable $U$ is generated as
\begin{IEEEeqnarray}{c}
U=(\Delta_{1,1},\ldots,\Delta_{N,I},V_{1,1},\ldots,V_{L,S},\yq{\Lambda_{1,1},\ldots,\Lambda_{I,S}})\sim \textrm{Unif} \big\{\mathbb{F}_q^{\frac{B(NIF+LS+IS)}{LF}}\big\},\notag
\end{IEEEeqnarray} 
where $\mbox{Unif}\{\cdot\}$ denotes the uniform distribution over the set in the braces.  
Similar to \eqref{eqn:partition:packets}, $\Delta_{n,i}$ is partitioned into $F$ equal-size random packets:
\begin{IEEEeqnarray}{rcll}
\Delta_{n,i}&=&(\Delta_{n,i,1},\ldots,\Delta_{n,i,F}),\quad& \forall\, n\in[N], i\in[I],\label{eqn:partition:random}
\end{IEEEeqnarray}
where $\Delta_{n,i,j}\in\mathbb{F}_q^{\frac{B}{LF}}$ is the $j$-th packet of the random subfile $\Delta_{n,i}$. 

Define two set of polynomials
\begin{IEEEeqnarray}{rCl}
\widetilde{W}_n(x)&=&\sum_{l=1}^{L}W_{n,l}x^{l-1}+\sum_{i=1}^{I}\Delta_{n,i}x^{L+i-1},~~\forall\, n\in[N],\IEEEeqnarraynumspace\label{eqn:polynomials}\\
\widetilde{V}_s(x)&=&\sum_{l=1}^{L}V_{l,s}x^{l-1}\yq{+\sum_{i=1}^I\Lambda_{i,s}x^{L+i-1}},\quad\forall\, s\in[S]. 
\end{IEEEeqnarray}
In accordance to the partitions in \eqref{eqn:partition:packets} and \eqref{eqn:partition:random},  each of the polynomials in \eqref{eqn:polynomials} can be decomposed into $F$ components, i.e., 
\begin{IEEEeqnarray}{rCl}
\widetilde{W}_{n}(x)&=&\big(\widetilde{W}_{n,1}(x),\ldots,\widetilde{W}_{n,F}(x)\big),~\forall\,n\in[N]\label{eqn:polynomial:partition}
\end{IEEEeqnarray}
where
\begin{IEEEeqnarray}{c}
\widetilde{W}_{n,j}(x)=\sum_{l=1}^{L}W_{n,l,j}x^{l-1}+\sum_{i=1}^{I}\Delta_{n,i,j}x^{L+i-1},~
\forall\,j\in[F].\IEEEeqnarraynumspace \label{eqn:polynomials:subWV}
\end{IEEEeqnarray} 
In the following, for a given $\sd{b}=(b_1,\ldots,b_N)\in\mathbb{F}_q^N$, we denote the linear combinations of $\{\widetilde{W}_{n}(x):n\in[N]\}$ and $\{\widetilde{W}_{n,j}(x):n\in[N]\}$  with coefficient vector $\sd{b}$ by
\begin{IEEEeqnarray}{rCl}
\widetilde{W}_{\sd{b}}(x)&:=&\sum_{n\in[N]}b_n\cdot\widetilde{W}_n(x),\notag\\
\widetilde{W}_{\sd{b},j}(x)&:=&\sum_{n\in[N]}b_n\cdot\widetilde{W}_{n,j}(x),\quad\forall\,j\in[F],\notag
\end{IEEEeqnarray} 
Notice that, by \eqref{eqn:polynomial:partition},  $\widetilde{W}_{\sd{b}}(x)$ is composed of $F$ components, i.e.,  $\widetilde{W}_{\sd{b}}(x)=\big(\widetilde{W}_{\sd{b},1}(x),\ldots,\widetilde{W}_{\sd{b},F}(x)\big)$. 

\paragraph*{Server Storage Design} Let $\alpha_1,\ldots,\alpha_{H}$ be $H$ distinct nonzero elements in $\mathbb{F}_q$. For each $h\in[H]$, the system evaluates the values of the polynomials $\widetilde{W}_n(x)$  and $\widetilde{V}_s(x)$ at $\alpha_h$, and stores them at server $h$, that is,  
\begin{IEEEeqnarray}{c}
Z_h=\big\{\widetilde{W}_{n}(\alpha_h):n\in[N]\big\}\cup\big\{\widetilde{V}_s(\alpha_h):s\in[S]\big\}.\label{eqn:PDA:Zh} 
\end{IEEEeqnarray}
Notice that, by \eqref{eqn:polynomial:partition},  each coded subfile $\widetilde{W}_{n}(\alpha_h)$ is composed of $F$ equal-size packets, which are the evaluations of the polynomials in \eqref{eqn:polynomials:subWV} at  $\alpha_h$:
\begin{IEEEeqnarray}{c}
\widetilde{W}_{n}(\alpha_h)=\big\{\widetilde{W}_{n,1}(\alpha_h),...,\widetilde{W}_{n,F}(\alpha_h)\big\},\quad\forall\,n\in[N].\notag
\end{IEEEeqnarray}

\paragraph*{Placement Phase} Each user~$k\in[K]$ locally generates a random vector $\sd{p}_k$ uniformaly over $\mathbb{F}_q^N$, and constructs its local cache $C_k$ as
\begin{subequations}\label{eqn:Zk}
\begin{IEEEeqnarray}{rCl}
C_k=&&\{\sd{p}_k\}\label{eqn:Zk:c}
\\
&&\cup\{W_{n,l,j}:n\in[N],l\in[L], j\in[F],a_{j,k}=*\}\IEEEeqnarraynumspace\label{eqn:Zk:a}\\
&&\cup\{W_{\sd{p}_k,l,j}+V_{l,a_{j,k}}:l\in[L],j\in[F],a_{j,k}\neq *\}.\IEEEeqnarraynumspace\label{eqn:Zk:b}
\end{IEEEeqnarray}
\end{subequations}

\paragraph*{Delivery Phase} Assume that user~$k\in[K]$ demands 
$W_{\sd{d}_k}$, for some $\sd{d}_k\in\mathbb{F}_q^N$.  Then user~$k\in[K]$ sends query $\sd{q}_k=\sd{d}_k+\sd{p}_k$ to all the servers, i.e., the queries $Q_{k,[H]}$ are constructed as 
\begin{IEEEeqnarray}{c}
Q_{k,h}=\sd{q}_k=\sd{d}_k+\sd{p}_k,\quad \forall\,h\in[H].\label{eqn:Qkh}
\end{IEEEeqnarray} 
Upon receiving the queries $Q_{[K],h}=\sd{q}_{[K]}$, each server $h\in[H]$ constructs $S$ coded signals, one for each $s\in[S]$:
\begin{IEEEeqnarray}{c}
\widetilde{Y}_{s}(\alpha_h):=\widetilde{V}_{s}(\alpha_h)+\sum_{\substack{(u,v)\in[F]\times[K]\\a_{u,v}=s}}\widetilde{W}_{\sd{q}_v,u}(\alpha_h).\notag
\end{IEEEeqnarray}
Then each server \yq{$h\in[H]\backslash\mathcal{A}$} sends the signal
\begin{IEEEeqnarray}{c}
X_h=\big(\sd{q}_{[K]},\widetilde{Y}_{[S]}(\alpha_h)\big)\label{eq:Xh}
\end{IEEEeqnarray}
to the users, \yq{and each server $h\in\mathcal{A}$ sends an arbitrary signal $\widetilde{X}_h$  that satisfies \eqref{XA:I}. }


\subsection{Verification of Constraint Conditions}\label{sec:scheme:verify}

In this subsection, we prove that, the constraints in \eqref{scheme:constraint}  are satisfied in the PDA based \yq{RSP-LFR-BA scheme. For clarity, we need the following lemma from coding theory: 
\begin{lemma}\label{lem:mds}(Capability of MDS Code \cite{LinShu}) An $(n,k)$ MDS code with dimension $k$ and length $n$ is capable of resisting $b$ adversarial errors and $u$ erasure errors if the  minimum distance $d_{\min}\triangleq n-k+1\geq 2b+u+1$. 
\end{lemma}}

\paragraph*{Server Security} Notice that, by the secret-sharing arguments \cite{Shamir1979}, for any $\mathcal{I}\subseteq[H]$ such that $|\mathcal{I}|=I$,
\begin{IEEEeqnarray}{c}
I\big(W_{[N]};\big\{\widetilde{W}_{[N]}(\alpha_h):h\in\mathcal{I}\big\}\big)=0.\label{eqn:secret-I}
\end{IEEEeqnarray} 
Therefore,
\begin{IEEEeqnarray}{rCl}
&&I\big(W_{[N]};Z_\mathcal{I}\big)\notag\\
&=&I\big(W_{[N]};\{\widetilde{W}_{[N]}(\alpha_h)\}_{h\in\mathcal{I}},\{V_{[S]}(\alpha_h)\}_{h\in\mathcal{I}}\big)\notag\\
&=&I\big(W_{[N]};\{\widetilde{W}_{[N]}(\alpha_h)\}_{h\in\mathcal{I}}\big)
+I\big(W_{[N]};\{V_{[S]}(\alpha_h)\}_{h\in\mathcal{I}}|\{\widetilde{W}_{[N]}(\alpha_h)\}_{h\in\mathcal{I}}\big)\notag\\
&=&0,\label{Izero}
\end{IEEEeqnarray}
where the last equality follows from \eqref{eqn:secret-I} and the fact that the random variables $V_{[L],[S]}$ \yq{and $\Lambda_{[I],[S]}$} are generated independently and uniformly over $\mathbb{F}_q^{\frac{B}{LF}}$, and independent of $W_{[N]}$ and $\{\widetilde{W}_{[N]}(\alpha_h)\}_{h\in\mathcal{I}}$, so that the random variables $\{V_{[S]}(\alpha_h)\}_{h\in\mathcal{I}}$ are distributed independently and  uniformly over $\mathbb{F}_q^{\frac{B}{LF}}$ and independent of $\big\{W_{[N]},\{\widetilde{W}_{[N]}(\alpha_h)\}_{h\in\mathcal{I}}\big\}$. 
\paragraph*{Robust Recovery} \yq{Notice that, by \eqref{eqn:polynomials} and \eqref{eqn:PDA:Zh}, for each $n\in[N]$, the coded subfiles stored across the servers form a $(H,I+L)$ Maximum Distance Separable (MDS) code:
\begin{IEEEeqnarray}{c}
\big(\widetilde{W}_{n}(\alpha_1),\ldots,\widetilde{W}_{n}(\alpha_h)\big)=\big(W_{n,1},\ldots,W_{n,L},\Delta_{n,1},\ldots,\Delta_{n,I}\big)\cdot\left[\begin{array}{cccc}
1&1&\ldots&1\\
\alpha_1&\alpha_2&\ldots&\alpha_H\\
\vdots&\vdots&\ddots&\vdots\\
\alpha_1^{L+I-1}&\alpha_2^{L+I-1}&\ldots&\alpha_H^{L+I-1}
\end{array}
\right]\IEEEeqnarraynumspace
\end{IEEEeqnarray}
By Lemma \ref{lem:mds}, the subfiles $W_{[N],[L]}$ and noises $\Delta_{[N],[I]}$ can be decoded with any contents at the servers in $\mathcal{J}$, if 
\begin{IEEEeqnarray}{c}
H-(I+L)+1\geq (H-J)+2A+1, \label{MDS:HIL}
\end{IEEEeqnarray}
even if the adversarial servers $\mathcal{J}\cap\mathcal{A}$ provides erroneous contents. In fact, by the definition of $L$ in \eqref{def:L},  \eqref{MDS:HIL} holds with equality.}

\paragraph*{Robust Decoding}

We need to show that for each user~$k\in[K]$ and any \yq{$\mathcal{J},\mathcal{A}\subseteq[H]$ such that $|\mathcal{J}|=J$, and $|\mathcal{A}|=A$, with the signals $X_{\mathcal{J}\backslash\mathcal{A}},\widetilde{X}_{\mathcal{J}\cap\mathcal{A}}$} and the cache content $C_k$, user~$k$ can decode its demanded scalar linear function $W_{\sd{d}_k}$, i.e., all the packets $W_{\sd{d}_k,[L],[F]}$.

For each $i\in[F]$ such that $a_{i,k}=*$, by~\eqref{eqn:Zk:a}, user~$k\in[K]$ has stored all the packets $W_{[N],[L],i}$, thus it can directly compute the packets $W_{\sd{d}_k,l,i}$ for each $l\in[L]$. 

Now, consider any $i\in[F]$ such that $a_{i,k}\neq *$. Fixed a subset $\mathcal{J}\subseteq[H]$ such that $|\mathcal{J}|=J$.  Let $s\triangleq a_{i,k}$,  consider the polynomial 
\begin{IEEEeqnarray}{rCl}
&&\widetilde{Y}_s(x)\notag\\
&=&\widetilde{V}_{s}(x)+\sum_{\substack{(u,v)\in[F]\times[K]\\a_{u,v}=s}}\widetilde{W}_{\sd{q}_v,u}(x)\notag\\
&=&\sum_{l=1}^LV_{l,s}x^{l-1}\yq{+\sum_{i=1}^I\Lambda_{i,s}x^{L+i-1}}+\sum_{\substack{(u,v)\in[F]\times[K]\\a_{u,v}=s}}\sum_{n=1}^Nq_{v,n}\cdot\widetilde{W}_{n,u}(x)\notag\\
&=&\sum_{l=1}^LV_{l,s}x^{l-1}+\sum_{i=1}^I\Lambda_{i,s}x^{L+i-1}+\sum_{\substack{(u,v)\in[F]\times[K]\\a_{u,v}=s}}\sum_{n=1}^Nq_{v,n}\cdot \bigg[\sum_{l=1}^LW_{n,l,u}x^{l-1}
+\sum_{i=1}^I\Delta_{n,i,u}x^{L+i-1}\bigg]\notag\\
&=&\sum_{l=1}^L\bigg(V_{l,s}+\sum_{\substack{(u,v)\in[F]\times[K]\\a_{u,v}=s}}W_{\sd{q}_v,l,u}\bigg)x^{l-1}+\sum_{i=1}^I\bigg(\Lambda_{i,s}+\sum_{\substack{(u,v)\in[F]\times[K]\\a_{u,v}=s}}\Delta_{\sd{q}_v,i,u}\bigg)x^{L+i-1}\notag\\
&=&\sum_{l=1}^LY_{l,s} x^{l-1}+\sum_{i=1}^I\widetilde{\Delta}_{i,s}x^{L+i-1},\label{poly:YtildeS}
\end{IEEEeqnarray}
where $Y_{l,s}$ and $\widetilde{J}_{i,s}$ are defined as 
\begin{subequations}\label{eqn:coefYJ}
\begin{IEEEeqnarray}{rCl}
Y_{l,s}&\triangleq& V_{l,s}+\sum_{\substack{(u,v)\in[F]\times[K]\\a_{u,v}=s}}W_{\sd{q}_v,l,u},\quad\forall\,l\in[L]\label{eqn:coefYJ:a}\\
\widetilde{\Delta}_{i,s}&\triangleq& \Lambda_{i,s}+\sum_{\substack{(u,v)\in[F]\times[K]\\a_{u,v}=s}}\Delta_{\sd{q}_v,i,u},\quad\forall\,i\in[I].\label{def:DeltaT}
\end{IEEEeqnarray}
\end{subequations}
Notice that, the signal $\widetilde{Y}_s(\alpha_h)$ is the value of $\widetilde{Y}_s(x)$ evaluated at the point $\alpha_h$. \yq{Moreover, $(\widetilde{Y}_s(\alpha_1),\ldots,\widetilde{Y}_s(\alpha_H))$  is a $(H,I+L)$ 
 Maximum Distance Separable (MDS) code by \eqref{poly:YtildeS}: 
\begin{IEEEeqnarray}{c}
(\widetilde{Y}_s(\alpha_1),\ldots,\widetilde{Y}_s(\alpha_H))=(Y_{1,s},\ldots,Y_{L,s},\widetilde{\Delta}_{1,s},\ldots,\widetilde{\Delta}_{I,s})\cdot\left[\begin{array}{cccc}
1&1&\ldots&1\\
\alpha_1&\alpha_2&\ldots&\alpha_H\\
\vdots&\vdots&\ddots&\vdots\\
\alpha_1^{I+L-1}&\alpha_2^{I+L-1}&\ldots&\alpha_H^{I+L-1}
\end{array}
\right].
\end{IEEEeqnarray} 
Similar to the $J$-Recovery arguments, by Lemma \ref{lem:mds} and the fact   that \eqref{MDS:HIL} holds with equality, the signals $Y_{[L],s}$ and $\widetilde{\Delta}_{[I],s}$ can be decoded from the signals of servers $\mathcal{J}$ even if the adversarial servers send  erroneous signals.    
Furthermore, }
since $a_{i,k}=s$, for each $l\in[L]$, the signal $Y_{l,s}$ in~\eqref{eqn:coefYJ:a} can be written as
\begin{subequations}
\begin{IEEEeqnarray}{rCl}
Y_{l,s}
&=&V_{l,s}+W_{\sd{q}_k,l,i}+\sum_{\substack{(u,v)\in[F]\times[K]\\a_{u,v}=s,(u,v)\neq (i,k)}}W_{\sd{q}_v,l,u}\label{eqn:Yls:a}\\
&=&W_{\sd{d}_k,l,i}+(V_{l,a_{i,k}}+W_{\sd{p}_k,l,i})
+\sum_{\substack{(u,v)\in[F]\times[K]\\a_{u,v}=s=a_{i,k},(u,v)\neq(i, k)}}W_{\sd{q}_v,l,u},\label{eqn:Yls:b}
\end{IEEEeqnarray}
\end{subequations}
where \eqref{eqn:Yls:b} follows from $\sd{q}_k=\sd{p}_k+\sd{d}_k$. Therefore, user~$k\in[K]$ can decode $W_{\sd{d}_k,l,i}$ from the the signal $Y_{l,s}$ by canceling the remaining terms since 
\begin{enumerate}
\item the coded packet $V_{l,a_{i,k}}+W_{\sd{p}_k,l,i}$ is cached by user~$k$ by~\eqref{eqn:Zk:b};
\item for each $(u,v)\in[F]\times[K]$ such that $a_{u,v}=s$ and $(u,v)\neq (i,k)$, since $a_{i,k}=a_{u,v}=s$, by the definition of PDA,  $i\neq u,v\neq k$ and $a_{i,v}=a_{u,k}=*$. Thus, user~$k\in[K]$ stores all the packets $W_{[N],[L],u}$. Hence, user~$k$ can compute $W_{\sd{q}_v,l,u}$ for each $l\in[L]$.  
\end{enumerate}
\paragraph*{Signal Security}\yq{
We have
\begin{subequations}
\begin{IEEEeqnarray}{rCl}
&&I(W_{[N]};X_{[H]\backslash\mathcal{A}},\widetilde{X}_\mathcal{A})\notag\\
&=&I\big(W_{[N]};\sd{q}_{[K]},\{\widetilde{Y}_{h,[S]}(\alpha_h)\}_{h\in[H]\backslash\mathcal{A}},\widetilde{X}_{\mathcal{A}}\big)\notag\\
&\leq&I\big(W_{[N]};\sd{q}_{[K]},Y_{[L],[S]},\widetilde{\Delta}_{[I],[S]},\{\widetilde{Y}_{h,[S]}(\alpha_h)\}_{h\in[H]\backslash\mathcal{A}},\widetilde{X}_{\mathcal{A}},Z_\mathcal{A}\big)\notag\\
&=&I\big(W_{[N]};\sd{q}_{[K]},Y_{[L],[S]},\widetilde{\Delta}_{[I],[S]},\widetilde{X}_{\mathcal{A}},Z_\mathcal{A}\big)\label{eqn:exp:d}\\
&=&I\big(W_{[N]};\sd{q}_{[K]},Y_{[L],[S]},\widetilde{X}_{\mathcal{A}},Z_\mathcal{A}\big)\label{eqn:exp:d1}\\
&=&I\big(W_{[N]};\sd{q}_{[K]},\widetilde{X}_{\mathcal{A}},Z_\mathcal{A}\big)\label{eqn:exp:d2}\\
&=&I\big(W_{[N]};Z_{\mathcal{A}})+I(W_{[N]};\sd{q}_{[K]}|Z_\mathcal{A})+I(W_{[N]};\widetilde{X}_{\mathcal{A}}|Z_{\mathcal{A}},\sd{q}_{[K]})\notag\\
&=&0,\label{eqn:exp:f}
\end{IEEEeqnarray}
\end{subequations}
where 
\begin{itemize}
\item the equality
\eqref{eqn:exp:d} holds because $\{\widetilde{Y}_{h,[S]}(\alpha_h)\}_{h\in[H]\backslash\mathcal{A}}$ can be evaluated from $\sd{q}_{[K]},Y_{[L],[S]},\widetilde{\Delta}_{[I],[S]}$;
\item the equality \eqref{eqn:exp:d1} holds since $\widetilde{\Delta}_{[I],[S]}$ are  independent of $W_{[N]}$ by \eqref{def:DeltaT}; 
\item the equality \eqref{eqn:exp:d2} holds, because by the secret-sharing arguments \cite{Shamir1979}, the uniformly distributed keys $V_{[L],[S]}$ are independent of the coded versions for any $\mathcal{A}\subset[H]$ such that $|A|\leq I$, i.e., 
\begin{IEEEeqnarray}{c}
I(V_{[L],s};\{\widetilde{V}_{s}(\alpha_h)\}_{h\in\mathcal{A}})=0,\quad \forall\,s\in[S],
\end{IEEEeqnarray}
and thus $V_{[L],[S]}$ are independent of $(Z_\mathcal{A},W_{[N]},\sd{q}_{[K]})$. Moreover, 
the signals $Y_{[L],[S]}$ are padded by the keys $V_{[L],[S]}$ by \eqref{eqn:coefYJ:a}, and hence $Y_{[L],[S]}$ are independent of $(Z_{\mathcal{A}},\sd{q}_{[K]},W_{[N]})$. Furthermore, $Y_{[L],[S]}$ are independent of $\widetilde{X}_{\mathcal{A}}$ since $Y_{[L],[S]}\rightarrow (Z_{\mathcal{A}},\sd{q}_{[K]})\rightarrow \widetilde{X}_\mathcal{A}$ is a Markov chain. As a result, the signals $Y_{[L],[S]}$ are independent of $(Z_{\mathcal{A}},\sd{q}_{[K]},W_{[N]},\widetilde{X}_{\mathcal{A}})$. 
\item   the equality \eqref{eqn:exp:f} holds because:
\begin{itemize}
\item $I(W_{[N]};Z_{\mathcal{A}})=0$ holds due to \eqref{Izero};     
\item $I(W_{[N]};\sd{q}_{[K]}|Z_{\mathcal{A}})=0$ is obvious by \eqref{eqn:Qkh};
\item  $I(W_{[N]};\widetilde{X}_{\mathcal{A}}|Z_{\mathcal{A}},\sd{q}_{[K]})=0$ follows from the assumption \eqref{XA:I}.
\end{itemize}
\end{itemize}
}

\paragraph*{Demand Privacy} 
Notice that,  for any $\mathcal{S}\subseteq[K]$,
\begin{subequations}\label{eqn:user:privacy:proof}
\begin{IEEEeqnarray}{rCl}
&&I(\sd{d}_{[K]\backslash\mathcal{S}};C_{\mathcal{S}},\sd{d}_{\mathcal{S}},Q_{[K],[H]},Z_{[H]}\,|\,W_{[N]})\notag\\
&=&I(\sd{d}_{[K]\backslash\mathcal{S}};C_{\mathcal{S}},\sd{d}_{\mathcal{S}},\sd{q}_{[K]},Z_{[H]}\,|\,W_{[N]})\label{eqn:pri:a}\\
&=&I(\sd{d}_{[K]\backslash\mathcal{S}};C_{\mathcal{S}},\sd{d}_{\mathcal{S}},\sd{q}_{\mathcal{S}},Z_{[H]}\,|\,W_{[N]})
+I(\sd{d}_{[K]\backslash\mathcal{S}};\sd{q}_{[K]\backslash\mathcal{S}}\,|\,C_{\mathcal{S}},\sd{d}_{\mathcal{S}},\sd{q}_{\mathcal{S}},Z_{[H]},W_{[N]})\\
&=&0,\label{eqn:pri:b}
\end{IEEEeqnarray}
\end{subequations}
where \eqref{eqn:pri:a} follows from  \eqref{eqn:Qkh}; and \eqref{eqn:pri:b}                     follows since  the demands $\sd{d}_{[K]\backslash\mathcal{S}}$ are independent of the random variables $C_{\mathcal{S}},\sd{d}_{\mathcal{S}},\sd{q}_{[K]},Z_{[H]},W_{[N]}$, and $\sd{q}_k=\sd{p}_{k}+\sd{d}_k$ for each $k\in[K]\backslash\mathcal{S}$ where the random variables $\sd{p}_{[K]\backslash\mathcal{S}}$ are independently and uniformly distributed over $\mathbb{F}_q^{N}$.


\subsection{Performance}\label{sec:performance}
By \eqref{eqn:partiton:subfile} and \eqref{eqn:partition:packets}, each file is split into $LF$ equal-size packets, each of length $\frac{B}{LF}$ symbols, thus the subpacketization is $LF$. Denote the achieved MSC triple by $(M_{\sd{A}},T_{\sd{A}},R_{\sd{A}})$, then
\paragraph*{Memory Size}
For each user~$k\in[K]$, by the cached content in~\eqref{eqn:Zk}, for each $i\in[F]$ such that $a_{i,k}=*$, there are $LN$ associated packets cached by the user, one from each file (see~\eqref{eqn:Zk:a}). For each $i\in[F]$ such that $a_{i,k}\neq *$, there are $L$ associated coded packet cached at the user (see~\eqref{eqn:Zk:b}). In addition, the $\sd{p}_{k}$ in~\eqref{eqn:Zk:c} can be stored with $N$ symbols.
 Recall that, each column of a $(K,F,Z,S)$ PDA has $Z$ $``*"$s and $F-Z$ ordinary symbols, thus, the achieved memory size is
\begin{IEEEeqnarray}{rCl}
M_\mathbf{A}&=&\inf_{B\in\mathbb{N}^+}\frac{1}{B}\Big((Z \cdot LN+(F-Z)\cdot L)\frac{B}{LF}+N\Big)\notag\\
&=&\frac{F+Z \ (N-1)}{F}.\notag
\end{IEEEeqnarray}
\paragraph*{Storage Size} By \eqref{eqn:PDA:Zh}, since each coded subfile $\widetilde{W}_n(\alpha_h)$ is of $\frac{B}{L}$ symbols, and each $\widetilde{V}_s(\alpha_h)$ is of $\frac{B}{LF}$ symbols, the storage size is 
\begin{IEEEeqnarray}{rCl}
T_{\sd{A}}=\frac{1}{B}\Big(\frac{B}{L}\cdot N+\frac{B}{LF}\cdot S\Big)=\frac{N}{L}+\frac{S}{LF}. \notag
\end{IEEEeqnarray}
\paragraph*{Communication Load}
By~\eqref{eq:Xh}, each server $h\in[H]$ creates $S$ coded packets $\widetilde{Y}_{h,[S]}(\alpha_h)$, each of $\frac{B}{LF}$ symbols, and the coefficient vectors  $\sd{q}_{[K]}$ can be sent in $KN$ symbols, thus the achieved load is
\begin{IEEEeqnarray}{c}
\yq{R_{\mathbf{A}}=\inf_{B\in\mathbb{N}^+}\frac{1}{B}\Big( S \cdot\frac{ B}{LF}+ KN\Big)=\frac{S}{LF}.}\notag
\end{IEEEeqnarray}
\yq{
\begin{remark} Notice that by \eqref{eqn:Qkh}, each user $k\in[K]$ only needs to broadcast its query $\sd{q}_k$, a vector of length $N$, to all the servers. The queries  in  \eqref{eq:Xh} need to be conveyed to all users through the servers.     
 Alternatively, the users can share their queries directly to each other. 

It is worth pointing out that the cost for uploading the queries does not scale with the file size $B$, which means that it is igonrable compared to the downloading cost. As a result, it has no effect on the achieved performance of the system, even if the users share their queries directly. 
\end{remark}
}

\section{Lower Bounds of Storage Size and Communication Load}\label{sec:bound}
In this section, we prove Theorem \ref{thm:TRbounds}. In fact, \yq{without loss of generality, assume the adversarial servers $\mathcal{A}$ are not the first $I$ servers, i.e.,  $\mathcal{A}\cap [I]=\emptyset$, then 
\begin{subequations}
\begin{IEEEeqnarray}{rCl}
NB&=&H(W_{[N]})\notag\\
&=&I(W_{[N]}; Z_{[J]\backslash\mathcal{A}},\widetilde{Z}_{[J]\cap\mathcal{A}})+H(W_{[N]}|Z_{[J]\backslash\mathcal{A}},\widetilde{Z}_{[J]\cap\mathcal{A}})\notag\\
&=&I(W_{[N]}; Z_{[J]\backslash\mathcal{A}},\widetilde{Z}_{[J]\cap\mathcal{A}})\label{eqn:T-a}\\
&=&I(W_{[N]};Z_{[I]})+I(W_{[N]};Z_{[I+1:J]\backslash\mathcal{A}},\widetilde{Z}_{[I+1:J]\cap\mathcal{A}}|Z_{[I]})\notag\\
&=&I(W_{[N]};Z_{[I+1:J]\backslash\mathcal{A}},\widetilde{Z}_{[I+1:J]\cap\mathcal{A}}|Z_{[I]})\label{eqn:T-b}\\
&\leq&H(Z_{[I+1:J]\backslash\mathcal{A}},\widetilde{Z}_{[I+1:J]\cap\mathcal{A}}|Z_{[I]})\notag\\
&\leq&(J-I)T^*(M)B,\\
&=&(L+2A)T^*(M)B, \label{eqn:T-c}
\end{IEEEeqnarray}
\end{subequations}}
where \eqref{eqn:T-a} and \eqref{eqn:T-b} follows from the constraints \eqref{eqn:server-constraint:a} and \eqref{eqn:server-constraint:b}, respectively. Thus, the bound  \eqref{T:low} is proved.

Moreover, consider the case where each user demands a file, and denote $D_k$ the index of the file demanded by user $k$ for all $k\in[K]$. Denote the signal of server $h\in[H]$ under the demands $(D_1,\ldots,D_K)=(d_1,\ldots,d_K)$ by $X_{(d_1,\ldots,d_K),h}$. For each $u\in[\min\{\lfloor\frac{N}{2}\rfloor,K\}]$, consider the first $u$ caches $C_1,\cdots,C_u$.  For each $r\in[\lfloor\frac{N}{u}\rfloor]$ and $h\in[H]$, denote
\begin{IEEEeqnarray}{c}
X_{r,h}\triangleq X_{((r-1)u+1,(r-1)u+2,\ldots,ru,1,1,\ldots,1),h}.\notag
\end{IEEEeqnarray}
\yq{Let $\mathcal{J}\subseteq[H]$ be a subset of cardinality $J$. Let $\mathcal{I}\subseteq\mathcal{J}\backslash\mathcal{A}$ be a subset of cardinality $I$. }
Denote $\widetilde{N}_u=u\big\lfloor \frac{N}{u}\rfloor$, then the files $\{W_{n}:n=1,2,\ldots,\widetilde{N}_u\}$ can be decoded with the signals $\{X_{r,h}:r\in[\lfloor \frac{N}{u}\rfloor],h\in\mathcal{J}\}$ and the caches $\{C_1,\ldots,C_u\}$. Notice that $\widetilde{N}_u\geq u(\frac{N}{u}-\frac{u-1}{u})=N-u+1$, therefore
\begin{subequations}
\begin{IEEEeqnarray}{rl}
&\lefteqn{(N-u+1)B}\notag\\
&\leq  \widetilde{N}_u B\notag\\
&=I(W_{[\widetilde{N}_u]};X_{[\lfloor\frac{N}{u}\rfloor],\mathcal{J}},C_{[u]})\notag\\
&=I(W_{[\widetilde{N}_u]};X_{[\lfloor\frac{N}{u}\rfloor],\mathcal{I}})+I(W_{[\widetilde{N}_u]};X_{[\lfloor\frac{N}{u}\rfloor],\mathcal{J}\backslash\mathcal{I}},C_{[u]}\,|\,X_{[\lfloor\frac{N}{u}\rfloor],\mathcal{I}})\notag\\
&\leq I(W_{[N]};X_{[\lfloor\frac{N}{u}\rfloor],\mathcal{I}},Q_{[K],\mathcal{I}},Z_{\mathcal{I}})
+\sum_{h\in\mathcal{J}\backslash\mathcal{I}}\sum_{r=1}^{\lfloor \frac{N}{u}\rfloor}H(X_{r,h})+\sum_{k=1}^{u}H(C_u)\notag\\
&\leq I(W_{[N]};Q_{[K],\mathcal{I}},Z_{\mathcal{I}})+(J-I)\Big\lfloor \frac{N}{u}\Big\rfloor R^*(M)B+uMB\label{eqn:e:a}\\
&= I(W_{[N]};Z_{\mathcal{I}})+(J-I)\Big\lfloor \frac{N}{u}\Big\rfloor R^*(M)B+uMB\label{eqn:e:b}\\
&\leq (L+2A)\frac{N}{u} R^*(M)B+uMB,\label{eqn:e:c}
\end{IEEEeqnarray} 
\end{subequations}
where \eqref{eqn:e:a} follows since the signals $X_{[\lfloor\frac{N}{u}\rfloor],\mathcal{I}}$ are determined by $Z_{\mathcal{I}}$ and $Q_{[K],\mathcal{I}}$ by \eqref{eqn:XhPhi}; \eqref{eqn:e:b} follows since  the queries $Q_{[K],\mathcal{I}}$ are independent of $(W_{[N]},Z_{\mathcal{I}})$ by \eqref{eqn:query}, and \eqref{eqn:e:c} follows from the $I$-Security constraint \eqref{eqn:server-constraint:a} and the definition of $L$ in \eqref{def:L}. 

Therefore, 
\yq{
\begin{IEEEeqnarray}{rCl}
R^*(M)
&\geq&\frac{1}{I+2A}\cdot\frac{u(N-u+1-uM)}{N},
\quad\forall\, u\in\Big[\min\Big\{\Big\lfloor \frac{N}{2}\Big\rfloor,K\Big\}\Big],\label{RMLbound}
\end{IEEEeqnarray}}
 Then the lower bound \eqref{lowerboud:R} is proved.

\section{Optimality of MAN-PDA based RSP-LFR-BA Scheme}\label{sec:MAN:PDA}

Let $r(M)$ be the lower convex envelope of the following points:
\begin{IEEEeqnarray}{c}
\Big\{\Big(1+\frac{t(N-1)}{K},\frac{K-t}{t+1}\Big):t=0,1,\ldots,K\Big\}.\notag
\end{IEEEeqnarray}
Notice that, $r(M)$ is the achievable communication load with memory size $M$ for the key superpostion scheme under signal security and demand privacy in the single server setup in \cite{Y:D:SP-LFR}. Moreover, $r(M)$ has the following relationship with $T(M)$ and $R(M)$:
\begin{IEEEeqnarray}{c}
T(M)=\frac{1}{L}\big(N+r(M)\big),\quad R(M)=\frac{1}{L}\cdot r(M).\label{eqn:rMR}
\end{IEEEeqnarray}
We will use the following upper bound for $r(M)$, which was proved in \cite[Lemma 4]{Y:D:SP-LFR}:
\begin{IEEEeqnarray}{c}
r(M)\leq \frac{N-M}{M-1},\quad\forall\, M\in(1,N]. \label{eqn:rM}
\end{IEEEeqnarray}

 In the following subsections, we separately derive the lower bounds in \eqref{thm:bound}.

\subsection{Upper bound for $\frac{T(M)}{T^*(M)}$} 
 
By \eqref{T:low} in Theorem \ref{thm:TRbounds}, 
\begin{enumerate}
\item for any $K$ and $M\geq 2$, 
\begin{subequations}
\begin{IEEEeqnarray}{rCl}
\frac{T(M)}{T^*(M)}&\leq& \frac{\frac{1}{L}(N+r(M))}{\frac{N}{L+2A}}\label{eq:Tup-a}\\
&=&\frac{L+2A}{L}\Big(1+\frac{1}{N}\cdot r(M)\Big)\notag\\
&=&\Big(1+\frac{2A}{L}\Big)\Big(1+\frac{1}{N}\cdot\frac{N-M}{M-1}\Big)\label{eq:Tup-b}\\
&\leq&\Big(1+\frac{2A}{L}\Big)\Big(1+\frac{N-2}{N}\Big)\label{eq:Tup-c}\\
&\leq &2\Big(1+\frac{2A}{L}\Big),\notag
\end{IEEEeqnarray}
\end{subequations}
where \eqref{eq:Tup-a} and \eqref{eq:Tup-b} follows from \eqref{T:low} and  \eqref{eqn:rM}, respectively, and \eqref{eq:Tup-c} follows from the fact $M\geq 2$. 
\item for $K\leq N$, since $r(M)$ is upper bounded by $K$,  
\begin{IEEEeqnarray}{rCl}
\frac{T(M)}{T^*(M)}
&\leq&\frac{(N+K)/L}{N/(L+2A)}\notag\\
&=&\Big(1+\frac{K}{N}\Big)\Big(1+\frac{2A}{L}\Big)\label{eqn:1KN}\\
&\leq&2\Big(1+\frac{2A}{L}\Big). \notag
\end{IEEEeqnarray}
\end{enumerate}
\subsection{Upper bound for $\frac{R(M)}{R^*(M)}$} 

Now for each $u\in\big[\min\{\lfloor\frac{N}{2}\rfloor,K\}\big]$, define 
\begin{IEEEeqnarray}{c}
L_u(M)\triangleq \frac{1}{L+2A}\cdot \frac{u(N-u+1-uM)}{N},\quad\forall\,M\in[0,N]. \notag
\end{IEEEeqnarray}
Then the bound \eqref{lowerboud:R} in Theorem \ref{thm:TRbounds} indicates that for each $u\in\big[\min\{\lfloor\frac{N}{2}\rfloor,K\}\big]$,
\begin{IEEEeqnarray}{c}
R^*(M)\geq L_u(M),\quad \forall\,M\in[1,N].\label{ineq:RL}
\end{IEEEeqnarray}
For any $M\in[0,N]$, define
\begin{IEEEeqnarray}{c}
f(M)\triangleq\frac{1}{4(L+2A)N}\cdot\frac{(N-M)(N+M+2)}{M+1}.\notag
\end{IEEEeqnarray}
Notice that $f(M)$ is convex in $M$. The following lemma shows that $f(M)$ is a lower bound of $R^*(M)$ on $\big[\max\{2,\frac{N-2K}{2K+1}\},N\big]$. 
\begin{lemma}\label{Lemma:Rf} For any $M\in[\max\{2,\frac{N-2K}{2K+1}\},N\big]$, 
\begin{IEEEeqnarray}{c}
R^*(M)\geq f(M). \label{ineq:Rf}
\end{IEEEeqnarray}
\end{lemma}
\begin{IEEEproof} Consider the interval $\big[\max\big\{\frac{N-2\lfloor N/2\rfloor}{2\lfloor N/2\rfloor+1},\frac{N-2K}{2K+1}\big\},N\big]$, which can be split into $\min\{\lfloor\frac{N}{2}\rfloor,K\}$ intervals:
\begin{IEEEeqnarray}{c}
\Big[\max\Big\{\frac{N-2\lfloor N/2\rfloor}{2\lfloor N/2\rfloor+1},\frac{N-2K}{2K+1}\Big\},N\Big]=
\bigcup_{u=1}^{\min\{\lfloor\frac{N}{2}\rfloor,K\}}\Big[\frac{N-2u}{2u+1},\frac{N-2u+2}{2u-1}\Big].\notag
\end{IEEEeqnarray}
Notice that, by the fact $\lfloor \frac{N}{2}\rfloor\geq\frac{N-1}{2}$, $\frac{N-2\lfloor N/2\rfloor}{2\lfloor N/2\rfloor+1}\leq\frac{1}{N}\le 2$, the interval $\big[\max\big\{\frac{N-2\lfloor N/2\rfloor}{2\lfloor N/2\rfloor+1},\frac{N-2K}{2K+1}\big\},N\big]$ encloses $[\max\{2,\frac{N-2K}{2K+1}\},N\big]$ as its sub-interval. Therefore, for any $M\in[\max\{2,\frac{N-2K}{2K+1}\},N\big]$, there exists $u\in[\min\{\lfloor\frac{N}{2}\rfloor,K\}]$ such that
$M\in\big[\frac{N-2u}{2u+1},\frac{N-2u+2}{2u-1}\big]$. It is easy to verify:
\begin{IEEEeqnarray}{rCl}
L_{u}\Big(\frac{N-2u}{2u+1}\Big)&=&f\Big(\frac{N-2u}{2u+1}\Big)\notag\\
&=&\frac{N+1}{(L+2A)N}\cdot\frac{u(u+1)}{2u+1},\notag\\
L_{u}\Big(\frac{N-2u+2}{2u-1}\Big)&=&f\Big(\frac{N-2u+2}{2u-1}\Big)\notag\\
&=&\frac{N+1}{(L+2A)N}\cdot\frac{u(u-1)}{2u-1}.\notag
\end{IEEEeqnarray}
That is $L_u(M)$ and $f(M)$ concides on the end points of the interval $\big[\frac{N-2u}{2u+1},\frac{N-2u+2}{2u-1}\big]$. Since $f(M)$ is convex on $M$, we concludes that 
\begin{IEEEeqnarray}{c}
L_u(x)\geq f(x),\quad\forall\,x\in\Big[\frac{N-2u}{2u+1},\frac{N-2u+2}{2u-1}\Big]. \label{ineq:Lf}
\end{IEEEeqnarray}
Therefore, with \eqref{ineq:RL} and \eqref{ineq:Lf}, we concludes \eqref{ineq:Rf}, 
which holds for all $M\in\big[\max\big\{2,\frac{N-2K}{2K+1}\big\},N\big]$.
\end{IEEEproof}
Therefore, 
\begin{enumerate}
\item for any $K$, $M\geq \max\big\{2,\frac{N-2K}{2K+1}\big\}$,  by \eqref{eqn:rMR}, \eqref{eqn:rM} and Lemma \ref{Lemma:Rf}, 
\begin{IEEEeqnarray}{rCl}
&&\frac{R^(M)}{R^*(M)}\notag\\
&\leq& \frac{\frac{1}{L} \cdot r(M)}{f(M)}\notag\\
&\leq&\frac{\frac{1}{L}\cdot\frac{N-M}{M-1}}{\frac{1}{4N(L+2A)}\cdot\frac{(N-M)(N+M+2)}{M+1}}\notag\\
&=&4N\cdot\Big(1+\frac{2A}{L}\Big)\cdot\frac{M+1}{(M-1)(N+M+2)}\notag\\
&=&4N\cdot\Big(1+\frac{2A}{L}\Big)\cdot\Big(\frac{1}{N+M+2}+\frac{2}{(M-1)(N+M+2)}\Big)\notag\\
&\leq&12\cdot\Big(1+\frac{2A}{L}\Big)\cdot \frac{N}{N+4}\label{ineq:M2}\\
&<&12\Big(1+\frac{2A}{L}\Big),\notag
\end{IEEEeqnarray}
where in \eqref{ineq:M2}, we used the fact $M\geq2$. This completes the proof of second inequalilty in \eqref{thm:bound}.
\item for $K\leq N$ and $ M\leq \max\big\{2,\frac{N-2K}{2K+1}\big\}$, 
\begin{IEEEeqnarray}{c}
R(M)\leq \frac{K}{L}, \label{eqn:upRM}
\end{IEEEeqnarray}
and since $R^*(M)$ must be non-increasing with $M$, 
\begin{IEEEeqnarray}{c}
R^*(M)\geq R^*\Big(\max\Big\{2,\frac{N-2K}{2K+1}\Big\}\Big)\label{eqn:lowRM}
\end{IEEEeqnarray}
\begin{enumerate}
\item if $N\geq 6K+2$, $\max\big\{2,\frac{N-2K}{2K+1}\big\}=\frac{N-2K}{2K+1}$, by \eqref{ineq:Rf}, 
\begin{IEEEeqnarray}{rCl}
&&R^*\Big(\max\Big\{2,\frac{N-2K}{2K+1}\Big\}\Big)\notag\\
&\geq& f\Big(\frac{N-2K}{2K+1}\Big)\notag\\
&=&\frac{1}{4(L+2A)N}\cdot\frac{\big(N-\frac{N-2K}{2K+1}\big)\big(N+\frac{N-2K}{2K+1}+2)}{\frac{N-2K}{2K+1}+1}\notag\\
&=&\frac{1}{4(L+2A)N}\cdot\frac{\big(N+1-\frac{N+1}{2K+1}\big)\big(N+1+\frac{N+1}{2K+1}\big)}{\frac{N+1}{2K+1}}\notag\\
&=&\frac{1}{4(L+2A)}\cdot\frac{N+1}{N}\cdot2K\cdot\Big(1+\frac{1}{2K+1}\Big)\notag\\
&\geq&\frac{K}{2(L+2A)}, \label{eqn:low2RM}
\end{IEEEeqnarray}
thus by \eqref{eqn:upRM}, \eqref{eqn:lowRM} and \eqref{eqn:low2RM}, 
\begin{IEEEeqnarray}{c}
\frac{R(M)}{R^*(M)}\leq 2\Big(1+\frac{2A}{L}\Big)\leq 12\Big(1+\frac{2A}{L}\Big). \label{eqn:2ratio}
\end{IEEEeqnarray}
\item if $N<6K+2$ and $N\geq 4$,  $\max\big\{2,\frac{N-2K}{2K+1}\big\}=2$, by \eqref{ineq:Rf}, 
\begin{IEEEeqnarray}{rCl}
&&R^*\Big(\max\Big\{2,\frac{N-2K}{2K+1}\Big\}\Big)\notag\\
&\geq& f(2)\notag\\
&=&\frac{1}{4(L+2A)N}\cdot\frac{(N-2)(N+4)}{3}. \label{eqn:low3RM}
\end{IEEEeqnarray}
thus by \eqref{eqn:upRM}, \eqref{eqn:lowRM} and \eqref{eqn:low3RM}, 
\begin{IEEEeqnarray}{rCl}
\frac{R(M)}{R^*(M)}&\leq&\Big(1+\frac{2A}{L}\Big)\cdot\frac{12KN}{N^2+2N-8}\notag\\
&\leq &\Big(1+\frac{2A}{L}\Big)\cdot\frac{12N^2}{N^2+2N-8}\notag\\
&\leq &12\Big(1+\frac{2A}{L}\Big),\notag
\end{IEEEeqnarray}
where the last step follows from $N\geq 4$. 
\item if $N<6K+2$ and $N\leq 3$, it is easy to obtain 
\begin{IEEEeqnarray}{c}
R(M)\leq \frac{1}{L}\cdot K\Big(1-\frac{M-1}{N-1}\Big),\notag
\end{IEEEeqnarray}
and by \eqref{lowerboud:R} in Theorem \ref{thm:TRbounds},
\begin{IEEEeqnarray}{c}
R^*(M)\geq \frac{1}{L+2A}\Big(1-\frac{M}{N}\Big). \notag
\end{IEEEeqnarray}
thus,
\begin{IEEEeqnarray}{rCl}
\frac{R(M)}{R^*(M)}&\leq&\frac{KN}{N-1}\Big(1+\frac{2A}{L}\Big)\\
&\leq& \frac{N^2}{N-1}\Big(1+\frac{2A}{L}\Big)\notag\\
&\leq& 4.5\Big(1+\frac{2A}{L}\Big)\notag\\
&\leq &12\Big(1+\frac{2A}{L}\Big). \notag
\end{IEEEeqnarray}
\end{enumerate}

\end{enumerate}
From the above arguments,  we conclude that, for all $K,N$  except for $K>N$ and $1\leq M\leq \max\big\{2,\frac{N-2K}{2K+1}\big\}$, 
\begin{IEEEeqnarray}{c}
\frac{R(M)}{R^*(M)}\leq 12\Big(1+\frac{2A}{L}\Big).\label{eqn:Rbound12}
\end{IEEEeqnarray}
 Notice that, when $K>N$, $\max\big\{2,\frac{N-2K}{2K+1}\big\} =2$. Hence, we proved  \eqref{eqn:Rbound12} for $K\leq N$ or $K>N,M\geq 2$.

\section{Conclusion}\label{sec:conclusion}
A  RSP-LFR-BA scheme was proposed within PDA framework, where the techniques of coded caching,  secret sharing,  key superpositons are integrated such that the signal security, users' demand privacy, and the blindness and robustness to colluding adversarial servers are simultaneously guaranteed. The storage size and the communication load  of MAN-PDA based RSP-LFR-BA scheme are shown to be to within a
multiplicative gap of at most $2(1+\frac{2A}{L})$ and $12(1+\frac{2A}{L})$  from optimal in all regimes, 
except for small memory regime with less files than users.

\end{document}